\documentclass[12pt]{article}
\usepackage{amssymb,cite}
\usepackage{amsmath}
\usepackage{epsfig}
\allowdisplaybreaks
\begin{document}
\title{\bf Viable Wormhole Solutions in Modified Gauss-Bonnet Gravity}
\author{M. Zeeshan Gul \thanks{mzeeshangul.math@gmail.com}
and M. Sharif \thanks {msharif.math@pu.edu.pk} \\
Department of Mathematics and Statistics, The University of
Lahore,\\
1-KM Defence Road Lahore-54000, Pakistan.}
\date{}
\maketitle
\begin{abstract}
In this paper, we use the embedding class-I technique to examine the
effect of charge on traversable wormhole geometry in the context of
$f(\mathcal{G})$ theory, where $\mathcal{G}$ is the Gauss-Bonnet
term. For this purpose, we consider static spherical spacetime with
anisotropic matter configuration to investigate the wormhole
geometry. The Karmarkar condition is used to develop a shape
function for the static wormhole structure. Using this developed
shape function, we construct a wormhole geometry that satisfies all
the required constraints and connects asymptotically flat regions of
the spacetime. To analyze the existence of traversable wormhole
geometry, we evaluate the behavior of energy conditions for various
models of this theory. This study reveals that viable traversable
wormhole solutions exist in this modified theory.
\end{abstract}

\section{Introduction}

The mysterious characteristics of our universe raise marvelous
questions for the research community. The presence of hypothetical
structures is assumed as the most controversial issue that yields
the wormhole (WH) structure. A WH is a hypothetical concept that
refers to a shortcut or a tunnel through spacetime. The basic idea
behind a WH is that it connects two separate points in spacetime,
allowing for faster-than-light travel. A WH that connects different
parts of the separate universe is called an inter-universe WH, but
an intra-universe WH joins distinct parts of the same universe. The
notion of a WH was first proposed in 1916 by the physicist Flamm
\cite{1}. He used the Schwarzschild solution to develop WH
structure. Later, Einstein and Rosen \cite{2} presented the concept
of Einstein-Rosen bridge, which examined that spacetime can be
connected by a tunnel-like structure, allowing for a shortcut or
bridge between two distinct regions of spacetime.

According to Einstein's theory of general relativity (GR), WHs can
exist if there is enough mass and energy to warp spacetime in a
specific way. They are also believed to be highly unstable and may
require exotic matter (which contradicts energy conditions) with
negative energy density to maintain their structure. However,
scientists continue to study the concept of WHs and their
implications for understanding the universe. According to Wheeler
\cite{3}, Schwarzschild WH solutions are not traversable due to the
presence of strong tidal forces at WH throat and the inability to
travel in two directions. Furthermore, the throat of WH rapidly
expands and then contracts, preventing access to anything. However,
it is analyzed that WHs would collapse immediately after the
formation \cite{4}. The possibility of a feasible WH is being
challenged due to the enormous amount of exotic matter. Thus, a
viable WH structure must have a minimum amount of exotic matter.
Morris and Thorne \cite{5} proposed the first traversable WH
solution.

The study of wormhole shape functions (WSFs) is one of the most
interesting subjects in traversable WH geometry. Shape functions
play a crucial role in determining the properties and behavior of
traversable WHs. They are mathematical functions that describe the
spatial geometry of a WH, specifically the throat's radius as a
function of the radial coordinate. By using different shape
functions, we can model various types of WHs with different
characteristics. The Morris-Thorne shape function is commonly used
to model spherical symmetric WHs \cite{6}. The choice of shape
function has a significant impact on WH properties, including
traversability, and the amount of exotic matter needed to keep the
WH throat open. Sharif and Fatima \cite{7} investigated non-static
conformal WHs using two different shape functions. Cataldo et al
\cite{8} studied static traversable WH solutions by constructing a
shape function that joins two non-flat regions of the universe.
Recently, many researchers \cite{9} proposed various shape functions
to describe the WH structure.

The WH geometry has been analyzed using several methods such as
solution of metric elements, constraints on fluid parameters and
specific type of the equation of state. Accordingly, the embedding
class-I method has been proposed that helps to examine the celestial
objects. One can embed $n$-dimensional manifold into
$(n+m)$-dimensional manifold according to this technique. The static
spherically symmetric solutions are examined using the embedding
class-I condition in \cite{10}. Karmarkar \cite{11} established a
necessary constraint for static spherical spacetime that belongs to
embedding class-I. Recently, spherical objects with different matter
distributions through the embedding class-I method have been studied
in \cite{12}-\cite{16}. The viable traversable WH geometry through
the Karmarkar constraint has been examined in \cite{17}. The effect
of charge can significantly impact the geometry of WHs. In
particular, the charge can create a repulsive force that pushes the
walls of the WH apart, making the throat of the WH wider. There has
been a lot of work exploring the influence of charge on the cosmic
structures \cite{18}-\cite{018d}. Sharif and Javed \cite{19}
investigated the impact of charge on thin-shell WH by employing the
cut-and-paste method.

General theory of relativity developed by Albert Einstein is the
most effective gravitational theory which explains a wide spectrum
of gravitational phenomena from small to large structures in the
universe. Gravitational waves have been confirmed by recent
observations and their power spectrum as well as properties are
consistent with those predicted by Einstein. In 1917, Einstein
introduced a term called the cosmological constant into his field
equations to account for the fact that the universe appeared to be
static and not expanding, which was the prevailing view at the time.
However, in 1929, Hubble's discovery of the expanding universe
prompted Einstein to remove the cosmological constant term from his
equations and revised them. In the late 1990s, different cosmic
observations reveal that our universe was in accelerated expansion
phase, which led physicists to revive the idea of a cosmological
constant \cite{20}. The problem, however, is that there is a large
difference between observed and predicted values of the cosmological
constant that explain the cosmic accelerated expansion. This
discrepancy is known as the cosmological constant problem. There are
also several other problems that keep the door open to extend GR.
Modifying GR is a fascinating approach to solve all of these
problems. This has led to the development of various extended
gravitational theories such as $f(\mathfrak{R})$ gravity \cite{21},
$f(\mathfrak{R},\mathcal{T})$ theory \cite{22},
$f(\mathcal{G},\mathcal{T})$ gravity \cite{23} and $f(Q)$ theory
\cite{24}. Sharif and Gul studied the Noether symmetry approach
\cite{23a}-\cite{023f}, stability of the Einstein universe
\cite{23b} and dynamics of gravitational collapse
\cite{23c}-\cite{0023g} in $f(\mathfrak{R},\mathcal{T}^{2})$ theory.

The Lovelock theory of gravity generalizes GR to higher dimensions.
It is named after mathematician David Lovelock, who developed the
theory in the 1970s. It is based on the idea that the gravitational
field can be described by a set of higher-order curvature tensors,
which are constructed from the Riemann tensor and its derivatives.
The key feature of Lovelock gravity is that it reduces to GR in four
dimensions while providing a more general description of gravity in
higher dimensions \cite{25}. One of the main applications of
Lovelock gravity is in the study of black holes in higher
dimensions. Lovelock gravity predicts the existence of black holes
with different properties than those predicted by GR. For example,
Lovelock gravity predicts that the event horizon of a black hole can
have a non-spherical shape in higher dimensions, which can have
important implications for the thermodynamics of black holes. Thus,
Lovelock gravity is an important theoretical framework for
understanding the behavior of gravity in higher dimensions, and it
has important implications for the study of black holes and other
astrophysical phenomena. The first Lovelock scalar is the Ricci
scalar $(\mathfrak{R})$, while the Gauss-Bonnet invariant
represented as
\begin{equation}\nonumber
\mathcal{G}=\mathfrak{R}_{\alpha\beta\lambda\delta}
\mathfrak{R}^{\alpha\beta\lambda\delta}+\mathfrak{R}^{2}
-4\mathfrak{R}_{\alpha\beta}\mathfrak{R}^{\alpha\beta},
\end{equation}
is the second Lovelock scalar \cite{26}. Here Ricci and Riemann
tensors are denoted by $\mathfrak{R}_{\alpha\beta}$ and
$\mathfrak{R}_{\alpha\beta\lambda\delta}$, respectively. Nojiri and
Odintsov \cite{27} established $f(\mathcal{G})$ gravity which
provides fascinating insights to the expansion of the universe at
present time. Moreover, this theory has no instability problems
\cite{28} and it is consistent with both solar system constraints
\cite{29} and cosmological structure \cite{30}.

The viable attributes of WHs provide fascinating outcomes in the
framework of modified gravitational theories. Lobo and Oliveira
\cite{31} used equations of state as well as various forms of shape
functions to examine the WH structures in $f(\mathfrak{R})$ theory.
Azizi \cite{32} analyzed the static spherically symmetric WH
solutions with particular equation of state in
$f(\mathfrak{R},\mathcal{T})$ framework. The traversable WH
structure in the background of $f(\mathcal{G})$ theory has been
analyzed in \cite{33}. Elizalde and Khurshudyan \cite{34} used the
barotropic equation of state to examine the viability and stability
of WH solutions in $f(\mathfrak{R},\mathcal{T})$ background. Sharif
and Hussain \cite{35} examined the viability and stability of static
spherical WH geometry in $f(\mathcal{G},\mathcal{T})$ gravity.
Mustafa and his collaborators \cite{36} studied compact spherical
structures with different considerations. We have considered static
spherically symmetric spacetime with Noether symmetry approach to
examine the WH solutions in $f(\mathfrak{R},\mathcal{T}^2)$ theory
\cite{38}. Godani \cite{39} studied the viable as well as stable WH
solutions in $f(\mathfrak{R},\mathcal{T})$ gravity. Malik et al
\cite{40} used embedding class-I technique to study the static
spherical solutions in $f(\mathfrak{R})$ theory. Recently, Sharif
and Fatima \cite{41} have employed the Karmarkar condition to
investigate the viable WH structures in
$f(\mathfrak{R},\mathcal{T})$ theory. Recently, the study of
observational constraints in modified $f(Q)$ gravity discussed in
\cite{041a} and thermal fluctuations of compact objects as charged
and uncharged BHs in $f(Q)$ gravity are explored in \cite{041b}.

This manuscript examines viable traversable WH solutions using the
embedding class-I technique in $f(\mathcal{G})$ theory. The behavior
of shape function and energy conditions is analyzed in this
perspective. We have arranged the paper in the following pattern. We
obtain WSF using Karmarkar condition in section \textbf{2}. In
section \textbf{3}, we construct the field equations in the
framework of $f(\mathcal{G})$ theory and examine the behavior of
energy conditions through different viable models of this theory.
The last section summarizes our outcomes.

\section{Karmarkar Condition and WH Geometry}

Here, we use embedding class-I technique to formulate the WSF that
determines the WH geometry. It is important to note that the
Karmarkar condition is just one approach among various methods used
to study WHs and gravitational solutions. It is worthwhile to
mention here that Karmarkar condition is a geometric condition that
involves the Riemann curvature tensor, which is a geometric quantity
characterizing the curvature of spacetime. The condition is
independent of any specific gravitational theory and is a set of
mathematical relationships that must be satisfied for a given
spacetime geometry. This condition is applied after the field
equations of a gravitational theory have been considered. Once a
solution to the field equations is found, the Karmarkar condition
ensures that this solution can be embedded consistently in a
higher-dimensional space. In this sense, it serves as a consistency
check on the obtained solution and provides a geometric perspective
on the allowed spacetime structures. A lot of work based on the
Karmarkar condition in the framework of different modified theories
has been done in \cite{41}-\cite{6a}. The main objective in
employing the Karmarkar condition is to find the solutions of metric
potentials used in the field equations of $f(G)$ theory. In adhering
to the Karmarkar condition, both metric potentials become apparent,
necessitating the assumption of one metric potential to deduce the
value of the other. It is imperative to note that metric potentials
involving Gauss-Bonnet terms cannot be considered in this context.

In this perspective, we consider static spherical metric as
\begin{equation}\label{1}
ds^{2}=-dt^{2}e^{a(r)}+dr^{2}e^{b(r)}+d\theta^{2}r^{2}
+d\phi^{2}\sin^{2}\theta.
\end{equation}
The non-vanish components of the Riemann curvature tensor with
respect to above spacetime are
\begin{eqnarray}\nonumber
\mathfrak{R}_{1212}&=&\frac{e^{a}(2a''+a'^{2}-a'b')}{4}, \quad
\mathfrak{R}_{3434}=\frac{r^{2}\sin^{2}\theta(e^{b}-1)}{e^{b}},
\\\nonumber
\mathfrak{R}_{1414}&=&\frac{r\sin^{2}\theta a'e^{a-b}}{2}, \quad
\mathfrak{R}_{2323}=\frac{rb'}{2}, \quad
\mathfrak{R}_{1334}=\mathfrak{R}_{1224}\sin^{2}\theta.
\end{eqnarray}
where $'=\frac{d}{dr}$. These Riemann components fulfill the
well-known Karmarkar condition as
\begin{eqnarray}\label{2}
\mathfrak{R}_{1414}&=&\frac{\mathfrak{R}_{1212}\mathfrak{R}_{3434}
+\mathfrak{R}_{1224}\mathfrak{R}_{1334}}{\mathfrak{R}_{2323}},\quad
\mathfrak{R}_{2323}\neq0.
\end{eqnarray}
Embedding class-I is the spacetime that satisfies the Karmarkar
condition. Solving this constraint, we obtain
\begin{equation}\label{2a}
\frac{a'b'}{1-e^{b}}=a'b'-2a''-a'^{2},
\end{equation}
where $e^{b}\neq1$. The corresponding solution is
\begin{equation}\label{3}
e^{b}=1+ \mu e^{a}a'^{2},
\end{equation}
where integration constant is denoted by $\mu$.

Now, we take the Morris-Thorne spacetime as \cite{5}
\begin{equation}\label{4}
ds^{2}=-dt^{2}e^{a(r)}+dr^{2}\frac{1}{1-\frac{\nu(r)}{r}}
+d\theta^{2}r^{2}+d\phi^{2}r^{2}\sin\theta.
\end{equation}
Here, shape function is denoted by $\nu(r)$ and the metric
coefficient $a(r)$ is defined as \cite{42}
\begin{equation}\label{5}
a(r)=\frac{-2c}{r},
\end{equation}
where the arbitrary constant is represented by $c$ and $a(r)$ is
called redshift function such that when $r\rightarrow\infty$,
$a(r)\rightarrow0$. Comparison of Eqs.(\ref{1}) and (\ref{4}) gives
\begin{equation}\label{6}
b(r)=\ln\left[\frac{r}{r-\nu(r)}\right].
\end{equation}
Using Eqs.(\ref{3}) and (\ref{6}), we have
\begin{equation}\label{7}
\nu(r)=r-\frac{r^{5}}{r^{4}+4c^{2}\mu e^{\frac{-2c}{r}}}.
\end{equation}
For a viable WH geometry, the given conditions must be satisfied
\cite{5}
\begin{enumerate}
\item
$\nu(r)<r$,
\item
$\nu(r)-r=0$ at $r=r_{0}$,
\item
$\frac{\nu(r)-r\nu'(r)}{\nu^{2}(r)}>0$ at $r=r_{0}$,
\item
$\nu'(r)<1$,
\item
$\frac{\nu(r)}{r}\rightarrow0$ when $r\rightarrow\infty$,
\end{enumerate}
where radius of WH throat is defined by $r_{0}$. At $r=r_{0}$, Eq.
(\ref{7}) gives trivial solution i.e., $\nu(r_{0})-r_{0}=0$.
Therefore, we redefine Eq.(\ref{7}) for non-trivial solution as
\begin{eqnarray}\label{8}
\nu(r)=r-\frac{r^{5}}{r^{4}+4c^{2}\mu e^{\frac{-2c}{r}}}+\eta.
\end{eqnarray}

For a viable WH geometry, the above conditions must be fulfilled.
These conditions are satisfied for $0<\eta<r_{0}$, otherwise the
required conditions are not satisfied and one cannot obtain the
viable WH structure. Using the condition (2) in the above equation,
we obtain
\begin{equation}\label{9}
\mu=\frac{r_{0}^{4}(r_{0}-\eta)}{4c^{2}e^{\frac{-2c^{2}}{r_{0}}}}.
\end{equation}
Inserting this value in Eq.(\ref{8}), it follows that
\begin{eqnarray}\label{10}
\nu(r)=r-\frac{r^{5}}{r^{4}+r_{0}^{4}(r_{0}-\eta)}+\eta.
\end{eqnarray}
Conditions (3) and (4) are also satisfied for the specified values
of $\eta$. Using condition (5) in Eq.(\ref{10}), we have
\begin{equation}\label{11}
\lim _{r\rightarrow\infty}\frac{\nu(r)}{r}=0.
\end{equation}
Thus, the formulated shape function gives asymptotically flat WH
geometry. We assume $r_{0}=2$, $c=-1$ and $\eta$=1.9 (blue), 1.8
(yellow), 1.7 (green), 1.6 (red) to analyze the graphical behavior
of the WSF. Figure \textbf{1} manifests that our developed shape
function through Karmarkar condition is physically viable as it
satisfies all the required conditions.
\begin{figure}
\epsfig{file=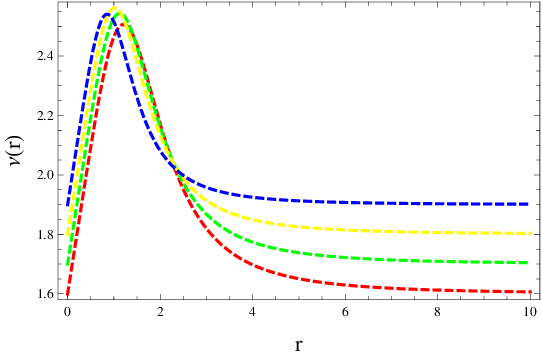,width=.5\linewidth}
\epsfig{file=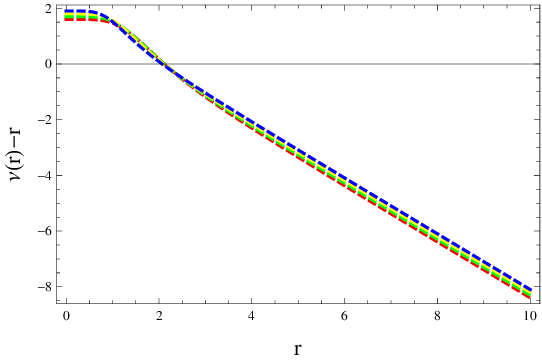,width=.5\linewidth}
\epsfig{file=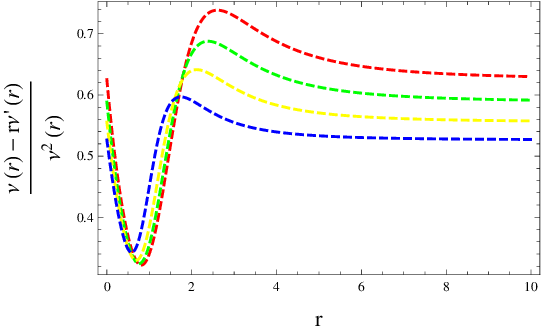,width=.5\linewidth}
\epsfig{file=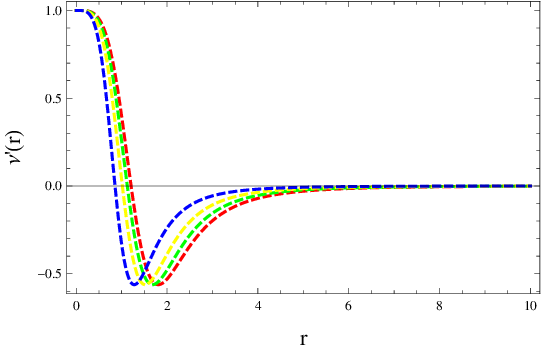,width=.5\linewidth}\center
\epsfig{file=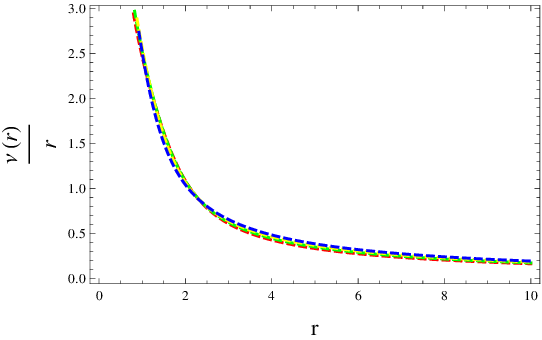,width=.5\linewidth}\caption{\label{F1}Behavior
of Morris-Thorne conditions corresponding to developed WSF.}
\end{figure}

\subsection{Embedding Diagram}

Classical geometry, specifically hyperbolic and elliptic
non-Euclidean geometry provides the foundation for embedding
theorems. A pseudo-sphere or an ordinary sphere can be visualized
within Euclidean three-space as these geometries have intrinsic
curvature. The development of embedding theorems has been greatly
influenced by Campbell's theorem in the framework of general
relativity (GR) \cite{42a}. The origin  of matter can be explained
by the resulting five-dimensional theory. In this five-dimensional
framework, the vacuum field equations yield the familiar Einstein
field equations supplemented with matter, leading to induced-matter
theory \cite{42b}. The incorporation of an additional dimension
serves the purpose of unification which greatly enhance our
understanding of physics in four dimensions. Moreover, the extra
dimension can be either timelike or spacelike. Consequently, the
resolution of particle-wave duality is achievable through the
utilization of five-dimensional dynamics, which exhibits two
distinct modes based on the nature of the additional dimensions such
as spacelike or timelike \cite{42c}. Thus, the theory of relativity
in five dimensions leads to unification of GR and quantum field
theory. The aforementioned analysis validates the effectiveness of
the mathematical model known as embedding theory, based on
Campbell's theorem. However, prior to employing this framework for
wormhole (WH) geometry, it is crucial to introduce a refinement in
this framework.

The induced-matter theory is a theoretical framework that extends
the concept of the Kaluza-Klein theory of gravity. It proposes a way
to incorporate matter into the theory that embeds four-dimensional
spacetime in a five-dimensional manifold, which is Ricci flat
\cite{42d}. This embedding process requires only one extra
dimension. In the context of embedding classes, one can embed
$n$-dimensional manifold into $(n+m)$-dimensional manifold. For
example, the interior Schwarzschild solution and the Friedmann
universe belong to class-I, while the exterior Schwarzschild
solution falls into class-II. Based on the similarities between WH
spacetime and the exterior Schwarzschild solution, it is assumed
that WH spacetime also belongs to class-II. Consequently, it can be
embedded in a six-dimensional flat spacetime. However, it is
noteworthy that a line element of class-II can be reduced to a line
element of class-I. This implies that the mathematical description
of WH spacetime can be transformed into a form that belongs to
embedding class-I. The mathematical model proposed by the
induced-matter theory has proven to be highly useful in the study of
cosmic objects, likely providing insights and explanations for
various phenomena and properties observed in the universe
\cite{42e}-\cite{42k}.

To extract useful information from WH geometry, we use embedding
diagram. It is an important tool for visualizing and understanding
the geometry of WHs and spacetime, in general. It allows us to
represent higher-dimensional curved spacetime in a lower-dimensional
Euclidean space. For a WH, a hypothetical tunnel connecting two
separate regions of spacetime, an embedding diagram can help us to
visualize the curvature and topology of the WH. By representing the
WH in a lower-dimensional space such as a two-dimensional plane, we
can gain insights into its shape and properties. To construct an
embedding diagram for a WH, we consider a spherically symmetric
spacetime. This means that the geometry of the WH remains the same
along spherical slices. In particular, we can focus on an equator
slice where the angular coordinate $\theta$ is fixed at
$\frac{\pi}{2}$. By choosing a constant time slice ($t=$ constant),
we can examine the spatial geometry of the WH independent of time.
In this equatorial slice, we can plot the spatial geometry of the WH
in the embedding diagram. The embedding diagram will typically be a
two-dimensional representation, where one axis represents the radial
distance from the WH center and the other axis represents some other
relevant coordinate or property. It is important to note that an
embedding diagram provides a simplified visualization and is not a
complete representation of the WH spacetime. Nevertheless, it can be
a helpful tool for gaining insights into the geometric properties of
WHs and understanding the effects of gravity in our universe.

Using these assumptions in Eq.(5), we have
\begin{equation}\label{11a}
ds^{2}=dr^{2}\left(\frac{r}{r-\nu}\right)+r^{2}d\phi^{2}.
\end{equation}
This embedding equation in cylindrical coordinates $(r,h,\phi)$ is
written as
\begin{equation}\label{11b}
ds^{2}=dr^{2}+dh^{2}+r^{2}d\phi^{2}=dr^{2}
\left[1+(\frac{dh}{dr})^{2}\right]+r^{2}d\phi^{2}.
\end{equation}
Comparison of Eqs.(\ref{11b}) and (\ref{11c}) gives
\begin{equation}\label{11c}
\frac{dh}{dr}=\pm\left(\frac{r}{\nu}-1\right)^{\frac{-1}{2}}.
\end{equation}
The embedding diagram for the upper universe ($h>0$) and the lower
universe ($h<0$) using a slice $t=$ constant and
$\theta=\frac{\pi}{2}$ corresponding to radial coordinate is shown
in Figures \textbf{2} and \textbf{3}, respectively. Moreover,
Eq.(\ref{11c}) indicates that the embedded surface is vertical at
the WH throat, i.e., $\frac{dh}{dr}\rightarrow\infty$. We also
examine that the space is asymptotically flat away from the throat
because $\frac{dh}{dr}$ tends to zero as $r$ tends to infinity. One
can visualize the upper universe for $h>0$ and the lower universe
$h<0$ in Figures \textbf{2} and \textbf{3}, respectively. One can
consider a $2\pi$ rotation around the $h$-axis for the full
visualization of the WH surface.
\begin{figure}
\epsfig{file=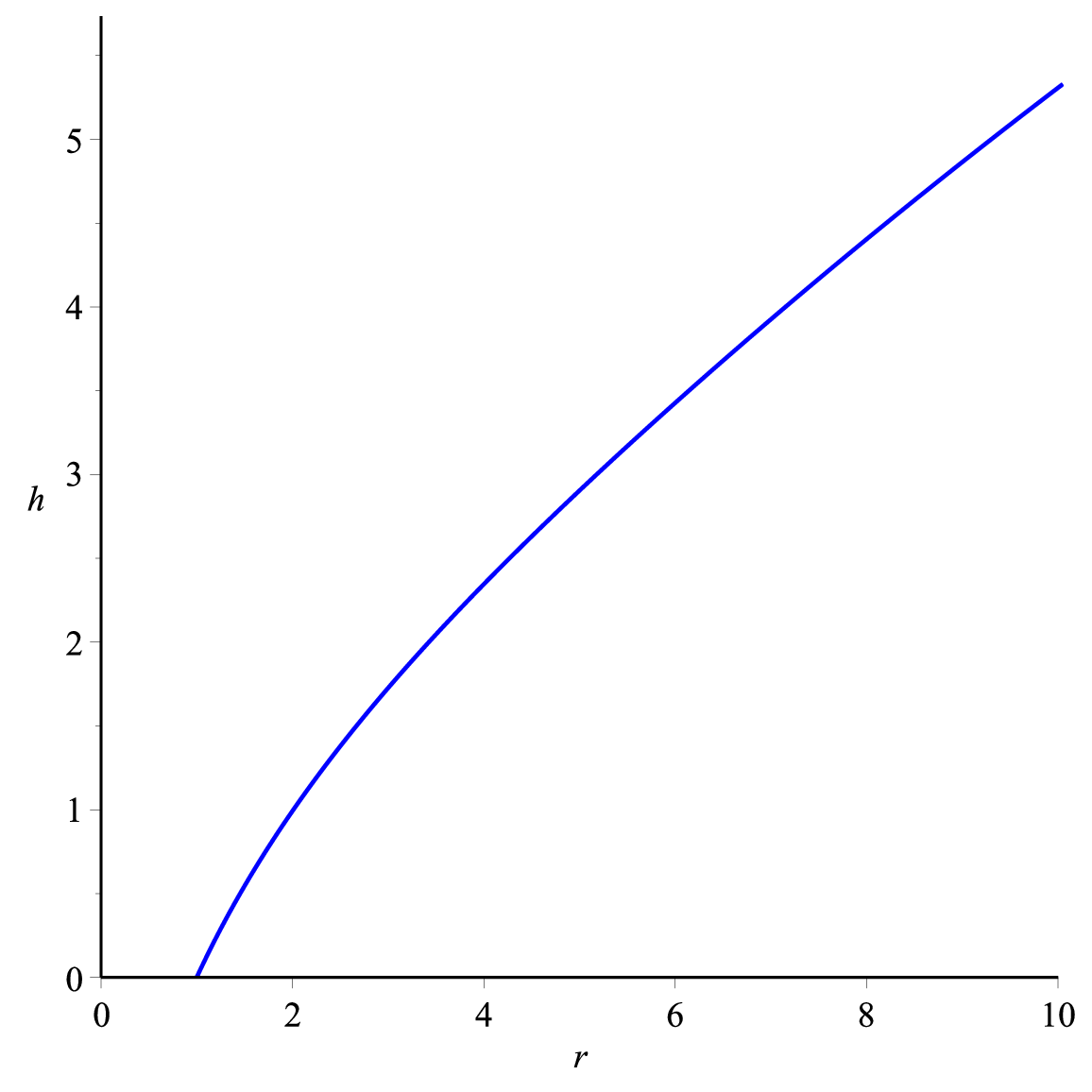,width=0.4\linewidth}
\epsfig{file=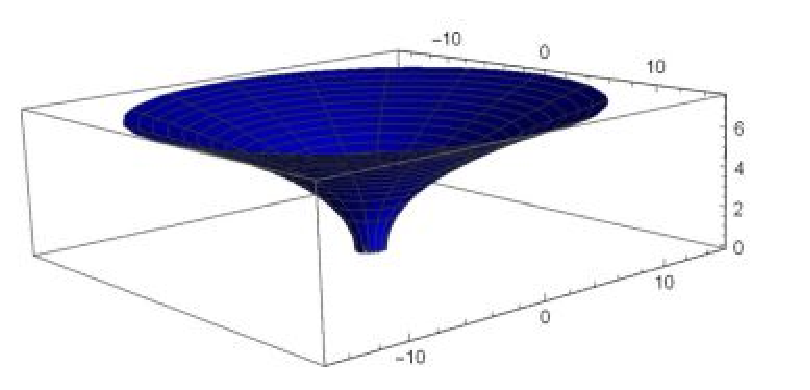,width=0.7\linewidth} \caption{Graph of
embedding diagram for $h(r)>0$ (upper universe) with slice $t=$
constant and $\theta=\frac{\pi}{2}$.}
\end{figure}
\begin{figure}
\epsfig{file=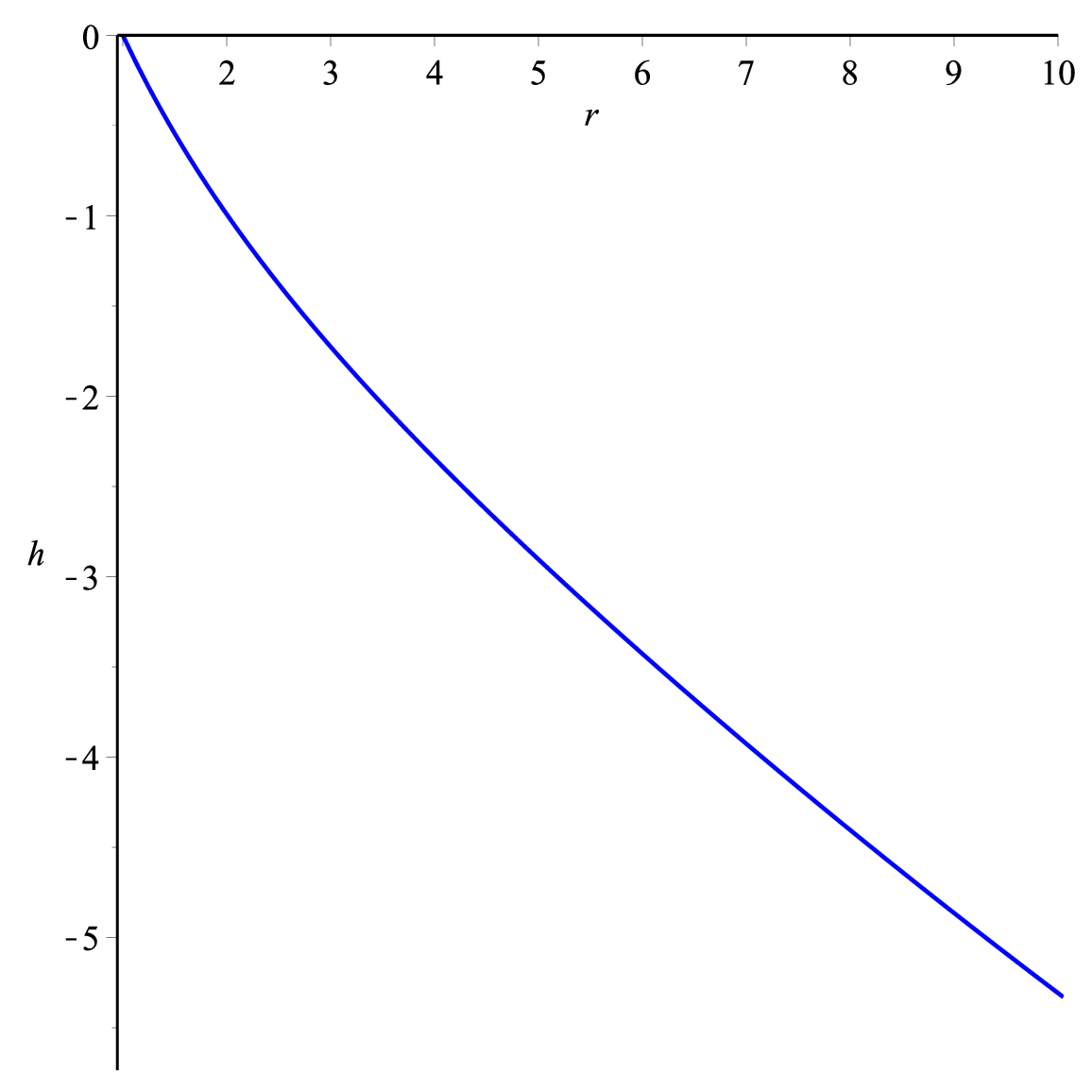,width=0.45\linewidth}
\epsfig{file=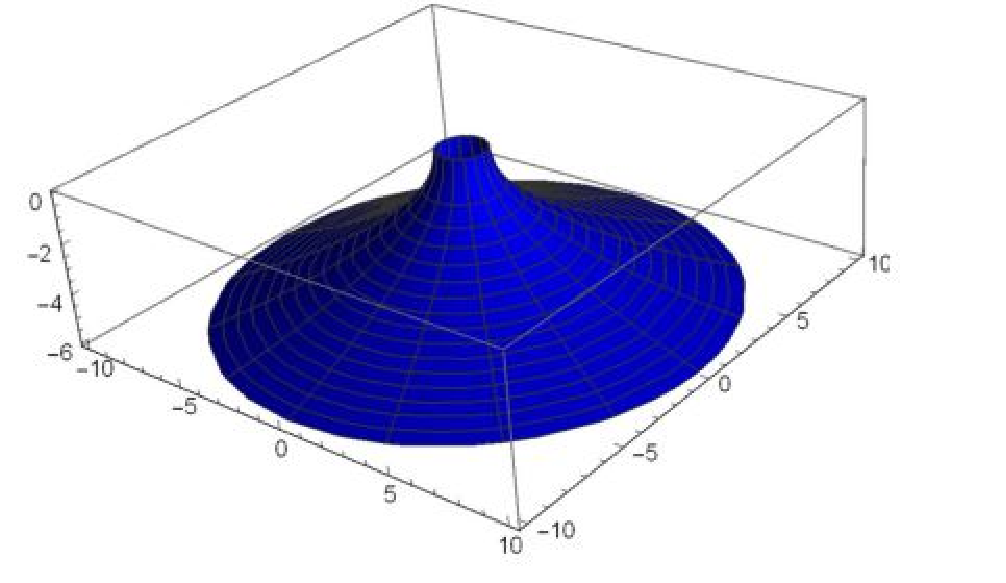,width=0.7\linewidth} \caption{Graph of
embedding diagram for $h(r)<0$ (lower universe) with slice $t=$
constant and $\theta=\frac{\pi}{2}$.}
\end{figure}

\section{Basic Formalism of $f(\mathcal{G})$ Gravity}

The corresponding action is defined as \cite{27}
\begin{equation}\label{12}
\mathcal{I}=\frac{1}{2\kappa}\int\mathfrak{R}{\sqrt{-\mathrm{g}}}d^4x
+\frac{1}{2\kappa}\int f (\mathcal{G}){\sqrt{-\mathrm{g}}}d^4x+\int
(\mathcal{L}_{m}+\mathcal{L}_{e})\sqrt{-\mathrm{g}}d^4x,
\end{equation}
where the matter-lagrangian and determinant of the line element are
represented by $\mathcal{L}_{m}$ and $\mathrm{g}$, respectively. The
electromagnetic-lagrangian is expressed as
\begin{eqnarray}\nonumber
\mathcal{L}_{e}=\epsilon\mathcal{F}_{\alpha\beta}\mathcal{F}^{\alpha\beta},
\quad \mathcal{F}_{\alpha
\beta}=\varphi_{\beta,\alpha}-\varphi_{\alpha,\beta},
\end{eqnarray}
where $\epsilon$ is an arbitrary constant and $\varphi_{\alpha}$ is
the four-potential. By varying Eq.(\ref{12}) corresponding to metric
tensor, we obtain
\begin{eqnarray}\nonumber
\mathfrak{R}_{\alpha\beta}-\frac{1}{2}\mathrm{g}_{\alpha\beta}
\mathfrak{R}&=&\kappa(\mathcal{T}_{\alpha\beta}
+\mathcal{E}_{\alpha\beta})-8\big(\mathfrak{R}_{\alpha\lambda
\beta\delta}+\mathfrak{R}_{\lambda\beta}\mathrm{g}_{\delta\alpha}
-\mathfrak{R}_{\lambda\delta}\mathrm{g}_{\alpha\beta}
-\mathfrak{R}_{\alpha\beta}\mathrm{g}_{\lambda\delta}
\\\nonumber
&+&\mathfrak{R}_{\alpha\delta}\mathrm{g}_{\beta\lambda}
+\frac{1}{2}\mathfrak{R}(g_{\alpha\beta}\mathrm{g}_{\lambda\delta}
-\mathrm{g}_{\alpha\delta}\mathrm{g}_{\beta\lambda})\big)
\nabla^{\lambda}\nabla^{
\delta}+\big(4\mathrm{g}_{\alpha\beta}\mathfrak{R}^{\lambda\delta}
\nabla_{\lambda}
\nabla_{\delta}
\\\nonumber
&+&2\mathfrak{R}\nabla_{\alpha}\nabla_{\beta}-2
g_{\alpha\beta}\mathfrak{R}\nabla_{\alpha}\nabla^{\alpha}
-4\mathfrak{R}^{\lambda}_{\alpha}\nabla_{\beta}\nabla_{\lambda}
-4\mathfrak{R}^{\lambda}_{\beta}\nabla_{\alpha}\nabla_{\lambda}
\\\label{13}
&-&4\mathfrak{R}_{\alpha\lambda\beta\delta}
\nabla^{\lambda}\nabla^{\delta}\big)f_{\mathcal{G}}.
\end{eqnarray}
Here, $f\equiv f(\mathcal{G})$ and $f_{\mathcal{G}}=\frac{\partial
f} {\partial \mathcal{G}}$. The stress-energy tensor of electric
field is expressed as
\begin{equation}\label{13a}
\mathcal{E}_{\alpha\beta}=\frac{1}{4\pi}\left[\frac{\mathcal{F}^{\lambda
\delta}\mathcal{F}_{\lambda\delta}
\mathrm{g}_{\alpha\beta}}{4}-\mathcal{F}^{\lambda}_{\alpha}
\mathcal{F}_{\beta\lambda}\right].
\end{equation}
We assume anisotropic matter distribution as
\begin{equation}\label{15}
\mathcal{T}_{\alpha\beta}=
\mathcal{V}_{\alpha}\mathcal{V}_{\beta}(\rho+p_{t})-p_{t}g_{\alpha\beta}
+\mathcal{U}_{\alpha}\mathcal{U}_{\beta}(p_{r}-p_{t}),
\end{equation}
where $\rho$ represents the energy density, $\mathcal{V}_{\alpha}$
is the four-velocity, $\mathcal{U}_{\alpha}$ defines four-vector,
$p_{r}$ denotes the radial pressure and $p_{t}$ is tangential
pressure.

The Maxwell field equations are defined as
\begin{eqnarray}\label{15a}
\mathcal{F}_{\alpha\beta;\lambda}=0, \quad F^{\alpha\beta}
_{;\beta}=4\pi \mathcal{J}^{\alpha},
\end{eqnarray}
where four-current is defined by $\mathcal{J}^{\alpha}$. In comoving
coordinates, $\varphi^{\alpha}$ and $\mathcal{J}^{\alpha}$ fulfill
the following relations
\begin{eqnarray}\label{15b}
\varphi^{\alpha}=\phi(r)\delta^{\alpha}_{0}, \quad
\mathcal{J}^{\alpha}=\sigma\mathcal{V}^{\alpha},
\end{eqnarray}
where $\sigma=\sigma(r)$ is the charge density. The resulting
electromagnetic field equation is
\begin{equation}\label{15c}
\varphi''+\bigg(\frac{2}{r}-\frac{a'}{2}-\frac{b'}{2}\bigg)\varphi'=4\pi
\sigma e^{\frac{a}{2}+b}.
\end{equation}
Integrating this equation, we get
\begin{eqnarray}\label{15d}
\varphi'=\frac{qe^{\frac{a+b}{2}}}{r^{2}}, \quad
q(r)=4\pi\int_{0}^{r}\sigma r^{2}e^{\frac{b}{2}}dr, \quad
E=\frac{q}{4\pi r^{2}},
\end{eqnarray}
where $E$ is the charge intensity and $q$ represents the charge
inside the interior of WH. Using Eqs.(\ref{1}) and (\ref{13}), we
obtain the field equations of charged anisotropic spherical system
as
\begin{eqnarray}\nonumber
\rho&=&\frac{e^{-2b}}{2r^{2}}\bigg[-2e^{b}+2e^{2b}-e^{2b}r^{2}f+e^{2b}r^{2}
\mathcal{G}f_{\mathcal{G}}+2b'(re^{b}-2(e^{b}-3)\mathcal{G}'f_{\mathcal{G}
\mathcal{G}})
\\\nonumber
&-&8(\mathcal{G}''f_{\mathcal{G}\mathcal{G}}+\mathcal{G}'^{2}f_{\mathcal{G}
\mathcal{G}\mathcal{G}})
(1-e^{b})-\frac{2q^{2}r^{2}}{8\pi r^{4}e^{-2b}}\bigg],
\\\nonumber
p_{r}&=&\frac{e^{-2b}}{2r^{2}}\bigg[e^{b}(2+e^{b}(r^{2}f-2))-e^{2b}r^{2}
\mathcal{G}f_{\mathcal{G}}+2a'(re^{b}-2(e^{b}-3)\mathcal{G}'f_{\mathcal{G}
\mathcal{G}})
\\\nonumber
&+&\frac{2q^{2}r^{2}}{8\pi r^{4}e^{-2b}}\bigg],
\\\nonumber
p_{t}&=&\frac{e^{-2b}}{4r}\bigg[-2e^{2b}r
\mathcal{G}f_{\mathcal{G}}+a'^{2}(re^{b}+4\mathcal{G}'f_{\mathcal{G}
\mathcal{G}})+2(e^{2b}rf
-b'e^{b}+(e^{b}r+4\mathcal{G}'f_{\mathcal{G}\mathcal{G}})a'')
\\\nonumber
&+&a'\big\{-b'(e^{b}r+12\mathcal{G}'f_{\mathcal{G}\mathcal{G}})
+2(e^{b}+4(\mathcal{G}''f_{\mathcal{G}\mathcal{G}}+\mathcal{G}'^{2}
f_{\mathcal{G}\mathcal{G}\mathcal{G}}))\big\}-\frac{2q^{2}r^{2}}{8\pi
r^{4}e^{-2b}}\bigg].
\end{eqnarray}
where
\begin{eqnarray}\nonumber
\mathcal{G}&=&-\frac{2}{r^{2}e^{2b}}\big[(e^{b}-3)a'b'-(e^{b}-1)
(2a''+a'^{2})\big],
\\\nonumber
\mathcal{G}'&=&\frac{2}{r^{3}e^{2b}}\bigg[2(e^{b}-1)a'^{2}+(6-e^{b})
ra'b'^{2} +2(e^{b}-1)(2a''-a''')+ra'(e^{b}-3)b''
\\\nonumber
&-&2(e^{b}-1)a''
+b'\{2(3-e^{b})a'+(3e^{b}-7)ra''+(e^{b}-2)ra'^{2}\}\bigg],
\\\nonumber
\mathcal{G}''&=&\frac{2}{r^{4}e^{2b}}\bigg[a'^{2}{6-6e^{b}
+(e^{b}-2)r^{2}b''}+(e^{b}-12)r^{2}a'b'^{3}-2\bigg\{a''
\big\{6(e^{b}-1)
\\\nonumber
&-&(2e^{b}-5)r^{2}b''+(e^{b}-1)r^{2}a''^{2}+(e^{b}-1)r(2a'''-4a''')
\big\}\bigg\}
+b'\bigg\{a'(6(e^{b}-3)
\\\nonumber
&+&4(e^{b}-2)r^{2}a''-3(e^{b}-6)r^{2}b'')
-4(e^{b}-2)ra'^{2}+r(a'''(5e^{b}-11)r
\\\nonumber
&-&4(3e^{b}-7)a'')\bigg\}
-rb'^{2}{4(e^{b}-5)ra''-(e^{b}-6)a'+(e^{b}-4)ra'^{2}} +ra'\big\{8a''
\\\nonumber
&\times&(e^{b}-1)-4(e^{b}-3)b''+r\big((e^{b}-3)b'''-2a'''(e^{b}-1)
\big)\big\}\bigg].
\end{eqnarray}

\section{Energy Conditions}

There are certain constraints named as energy conditions that must
be imposed on the matter to examine the presence of some viable
cosmic structures. These constraints are a set of inequalities that
impose limitations on the energy-momentum tensor which governs the
behavior of matter and energy in the presence of gravity. The
stress-energy tensor describes the distribution of energy, momentum
and stress in a given region of spacetime. There are several energy
conditions, each of which places different constraints on the
stress-energy tensor as
\begin{itemize}
\item Null energy constraint\\\\
This condition states that the energy density measured by any null
(light-like) observer is non-negative. The addition of energy
density and pressure components must be non-negative according to
this condition. Mathematically, this can be defined as
\begin{eqnarray}\nonumber
p_{r}+\rho\geq0, \quad p_{t}+\rho\geq0.
\end{eqnarray}
\item Dominant energy constraint\\\\
This determines that the energy flux measured by any observer cannot
exceed the energy density. Mathematically, it is expressed as
\begin{eqnarray}\nonumber
\rho-p_{r}\geq0, \quad \rho-p_{t}\geq0.
\end{eqnarray}
\item Weak energy constraint\\\\
The weak energy condition states that the energy density measured by
any observer is non-negative. Also, the sum of energy density and
pressure components are non-negative. Mathematically, this means
that
\begin{eqnarray}\nonumber
p_{r}+\rho\geq0,\quad p_{t}+\rho\geq0, \quad \rho\geq0.
\end{eqnarray}
\item Strong energy constraint\\\\
This energy condition is a stronger version of the weak energy
constraint and states that not only is the energy density
non-negative, but the addition of $\rho$ and pressure components is
also non-negative, defined as
\begin{eqnarray}\nonumber
p_{r}+\rho\geq0, \quad p_{t}+\rho\geq0, \quad
p_{r}+2p_{t}+\rho\geq0.
\end{eqnarray}
\end{itemize}
These energy bounds are significant in determining the presence of
cosmic structures. They also have implications for the behavior of
exotic matter and the existence of traversable WH geometry and other
hypothetical objects in spacetime. The viable WH structure must
violate these conditions.

The graphical behavior of energy conditions for $f(\mathcal{G})=0$
is given in Figure \textbf{4}. In the upper panel, the behavior of
$\rho+p_{t}$ is positive but negative behavior of $\rho+p_{r}$ shows
that the null energy condition is violated. Furthermore, the graphs
in the middle part manifest that the dominant energy condition is
violated as the behavior of $\rho-p_t$ is negative. The components
$\rho$ and $\rho+p_r+2p_t$ also exhibit negative trends, indicating
the violation of strong and weak energy conditions, respectively.
Thus, a viable traversable WH structure can be obtained in this
gravity model.
\begin{figure}
\epsfig{file=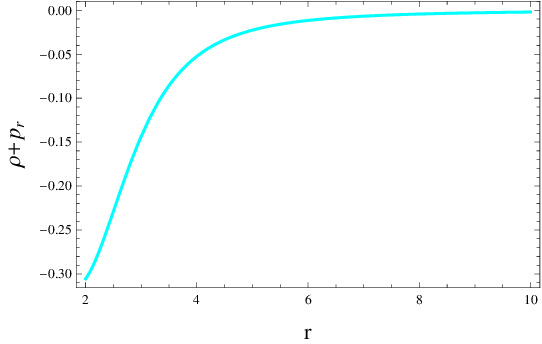,width=.5\linewidth}
\epsfig{file=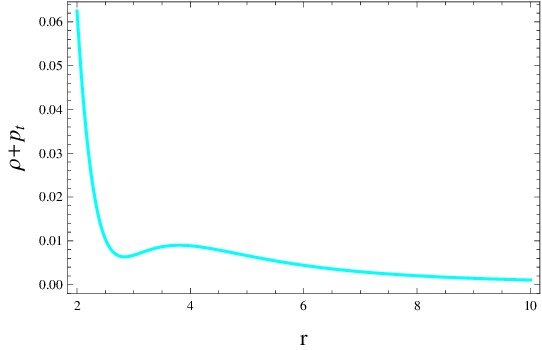,width=.5\linewidth}
\epsfig{file=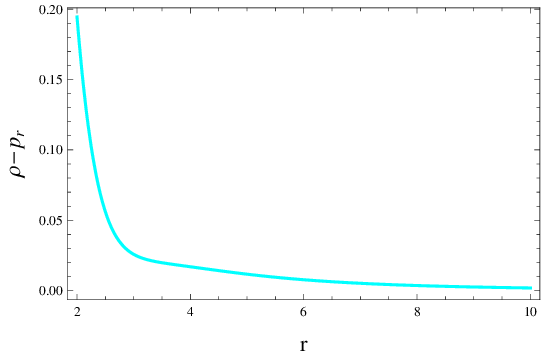,width=.5\linewidth}
\epsfig{file=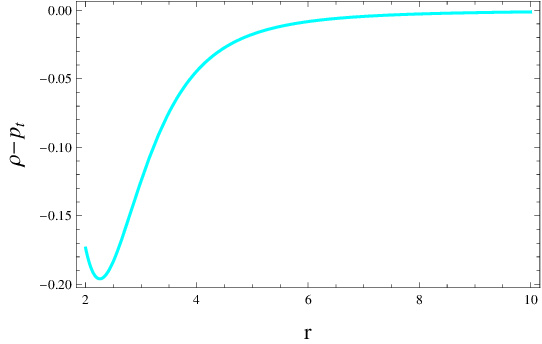,width=.5\linewidth}
\epsfig{file=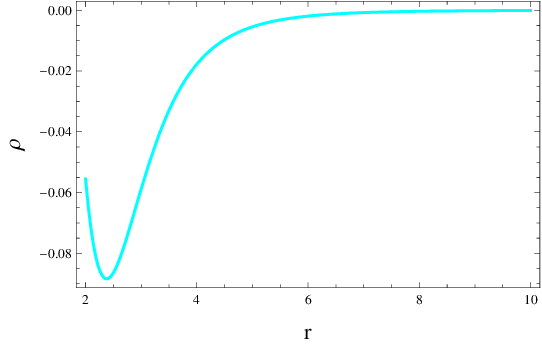,width=.5\linewidth}
\epsfig{file=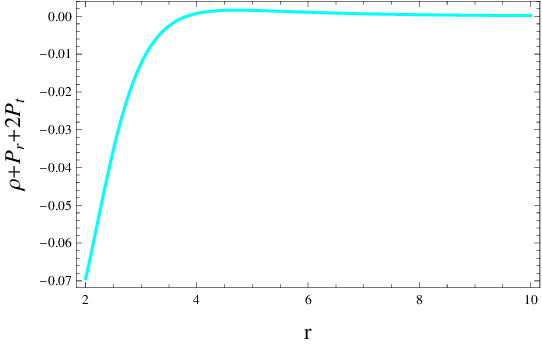,width=.5\linewidth}\caption{Plots of energy
conditions for $f(\mathcal{G})=0$.}
\end{figure}

\subsection{Viable $f(\mathcal{G})$ Models}

Here, we examine how various models of $f(\mathcal{G})$ affect the
geometry of WH. The outcomes of our research may uncover concealed
cosmological findings on both theoretical and astrophysical levels.
Some notable outcomes indicate the existence of additional
correction terms from modified gravitational theories, which could
yield some motivational results. These correction terms have a
significant influence on the collapse rate and the presence of
viable geometry as compared to GR. Hence, it is valuable to explore
alternative theories such as $f(\mathcal{G})$ to determine the
presence of hypothetical objects. This could serve as a mathematical
tool for examining various obscure features of gravitational
dynamics on a large scale. In the next subsections, we investigate
three distinct $f(\mathcal{G})$ models as

\subsection{Model l}

We first consider the power-law model with the logarithmic
correction term as \cite{43}
\begin{equation}\label{20}
f(\mathcal{G})=\gamma_{1}\mathcal{G}^{m_{1}}+\xi_{1}
\mathcal{G}\ln(\mathcal{G}),
\end{equation}
where $\gamma_{1}$, $\xi_{1}$ and $m_{1}$ are arbitrary constants.
Since this model allows extra degrees of freedom in the field
equations, therefore, it could provide observationally
well-consistent cosmic results. The resulting equations of motion
are
\begin{eqnarray}\nonumber
\rho&=&\frac{e^{-2b}}{2r^{2}}\bigg[-2e^{b}+2e^{2b}-e^{2b}r^{2}
(\gamma_{1}\mathcal{G}^{m_{1}}
+\xi_{1}\mathcal{G}\ln(\mathcal{G}))+e^{2b}r^{2}
\mathcal{G}(\gamma_{1}\mathcal{G}^{m_{1}-1}m_{1}
\\\nonumber
&+&\xi_{1}\ln(\mathcal{G})+\xi_{1})+2b'\big\{re^{b}-2(e^{b}-3)
\big\{\gamma_{1}
m_{1}(m_{1}-1)\mathcal{G}^{m_{1}-2}\mathcal{G}'+\xi_{1}
\mathcal{G}^{-1}\mathcal{G}'\big\}\big\}
\\\nonumber
&-&8(1-e^{b})\bigg\{\mathcal{G}''\big\{\gamma_{1}m_{1}(m_{1}-1)
\mathcal{G}^{m_{1}-2}+\xi_{1}\mathcal{G}^{-1}\big\}
+\mathcal{G}'^{2}\big\{\gamma_{1}(m_{1}-1)(m_{1}-2)
\\\nonumber
&\times&m_{1}\mathcal{G}^{m_{1}-3}-\xi_{1}\mathcal{G}^{-2}
\big\}\bigg\}-\frac{2q^{2}r^{2}}{8\pi
r^{4}e^{-2b}}\bigg],
\\\nonumber
p_{r}&=&\frac{e^{-2b}}{2r^{2}}\bigg[e^{b}(2+e^{b}(r^{2}(\gamma_{1}
\mathcal{G}^{m_{1}}+\xi_{1}\mathcal{G}\ln(\mathcal{G}))-2))-e^{2b}r^{2}
\mathcal{G}(\gamma_{1}\mathcal{G}^{m_{1}-1}m_{1}+\xi_{1}\ln(\mathcal{G})
\\\nonumber
&+&\xi_{1})+2a'\big\{re^{b}-2(e^{b}-3)\big\{\gamma_{1}m_{1}(m_{1}-1)
\mathcal{G}^{m_{1}-2}\mathcal{G}'+\xi_{1}\mathcal{G}^{-1}\mathcal{G}'
\big\}\big\}+\frac{2q^{2}r^{2}}{8\pi
r^{4}e^{-2b}}\bigg],
\\\nonumber
p_{t}&=&\frac{e^{-2b}}{4r}\bigg[-2e^{2b}r
\mathcal{G}(\gamma_{1}\mathcal{G}^{m_{1}-1}m_{1}+\xi_{1}\ln(\mathcal{G})
+\xi_{1})+a'^{2}\big\{re^{b}+4\big\{\gamma_{1}
m_{1}(m_{1}-1)
\\\nonumber
&\times&\mathcal{G}^{n-2}\mathcal{G}'+\xi_{1}\mathcal{G}^{-1}
\mathcal{G}'\big\}\big\}
+2\bigg\{e^{2b}r(\gamma_{1}\mathcal{G}^{m_{1}}+\xi_{1}
\mathcal{G}\ln(\mathcal{G}))
-b'e^{b}+\big\{e^{b}r+4\big\{\gamma_{1}m_{1}
\\\nonumber
&\times&(m_{1}-1)
\mathcal{G}^{m_{1}-2}\mathcal{G}'+\xi_{1}\mathcal{G}^{-1}
\mathcal{G}'\big\}\big\}
a''\bigg\}+a'\bigg[-b'(e^{b}r+12\big\{\gamma_{1} m_{1}(m_{1}-1)
\\\nonumber
&\times&\mathcal{G}^{n-2}\mathcal{G}'+\xi_{1}\mathcal{G}^{-1}
\mathcal{G}'\big\})
+2\bigg\{e^{b}+4\big\{\mathcal{G}''\big\{\gamma_{1}m_{1}(m_{1}-1)
\mathcal{G}^{m_{1}-2}+\xi_{1}\mathcal{G}^{-1}\big\}
\\\nonumber
&+&\mathcal{G}'^{2}\big\{\gamma_{1}
m_{1}(m_{1}-1)(m_{1}-2)\mathcal{G}^{m_{1}-3}-\xi_{1}
\mathcal{G}^{-2}\big\}\big\}\bigg\}\bigg]-\frac{2q^{2}r^{2}}{8\pi
r^{4}e^{-2b}}\bigg].
\end{eqnarray}
We consider radial dependent form of the charge as $q(r)=\chi r^{3}$
\cite{44}, where $\chi$ is an arbitrary constant. We choose
$\chi=0.0001$ for our convenience in all the graphs. Figures
\textbf{5} and \textbf{6} depict the graphical representation of
energy bounds for different values of model parameters. The behavior
of energy bounds for positive values of $\gamma_{1}$, $\xi_{1}$ and
$m_{1}$ is analyzed in Figure \textbf{5}. The plots in the upper
panel indicate that the behavior of $\rho+p_{r}$ and $\rho+p_{t}$ is
negative which implies that the null energy condition is violated.
The middle part shows that the dominant energy constraint is
violated due to the negative behavior of $\rho-p_{r}$ and
$\rho-p_{t}$. The behavior of energy density is also negative which
violates the weak energy condition. Although $\rho+p_{r}+2p_{t}$ is
positive near the center of the star but becomes negative at the
surface boundary, leading to a violation of the strong energy
condition.
\begin{figure}
\epsfig{file=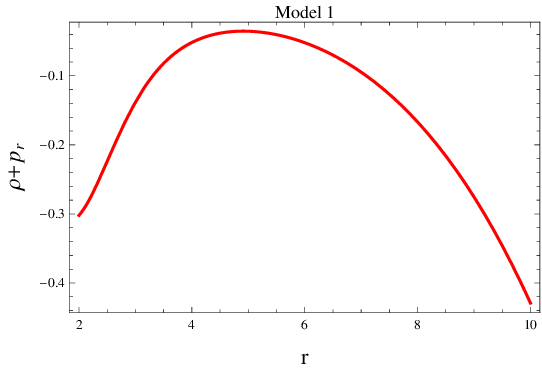,width=.5\linewidth}
\epsfig{file=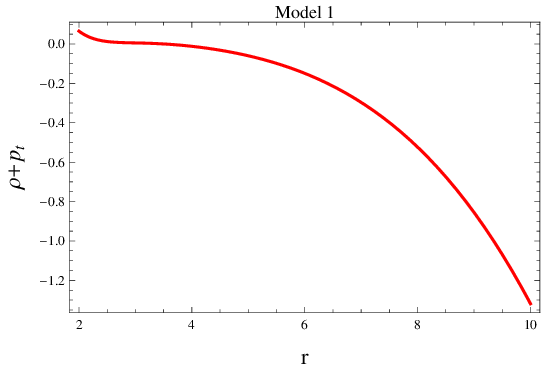,width=.5\linewidth}
\epsfig{file=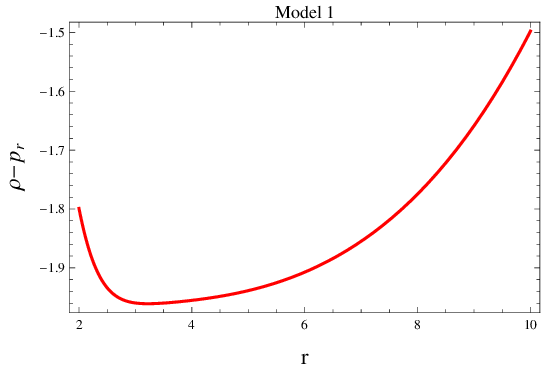,width=.5\linewidth}
\epsfig{file=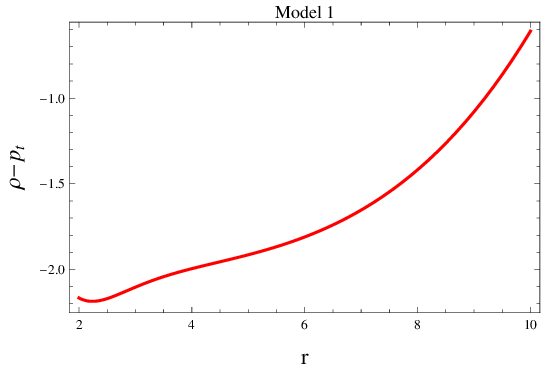,width=.5\linewidth}
\epsfig{file=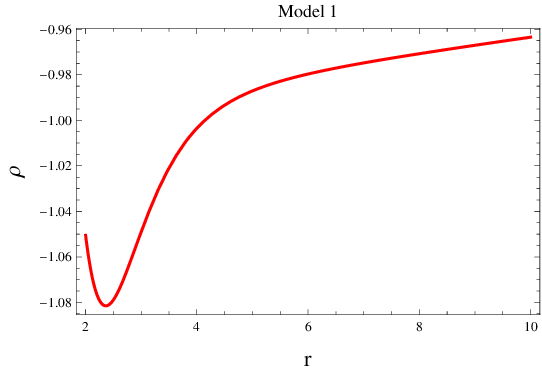,width=.5\linewidth}
\epsfig{file=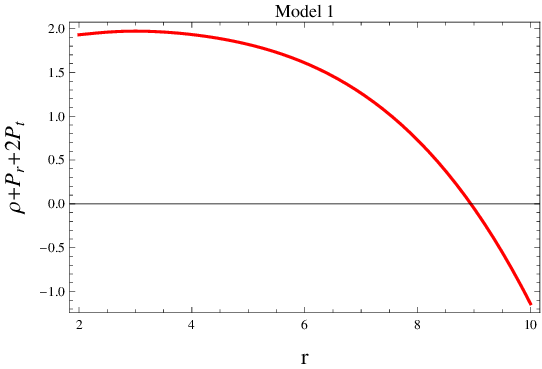,width=.5\linewidth}\caption{Plots of energy
conditions for $\gamma_{1}=2$, $\xi_{1}=0.0003$, $m_{1}=0.0005$.}
\end{figure}
\begin{figure}
\epsfig{file=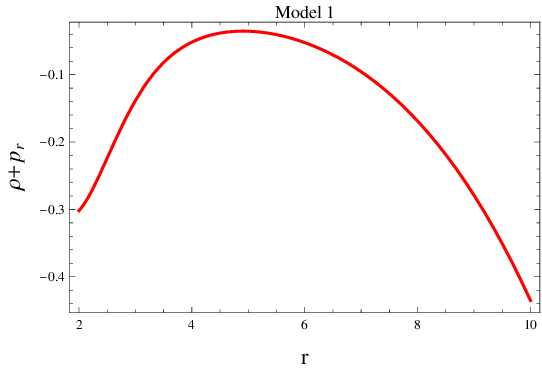,width=.5\linewidth}
\epsfig{file=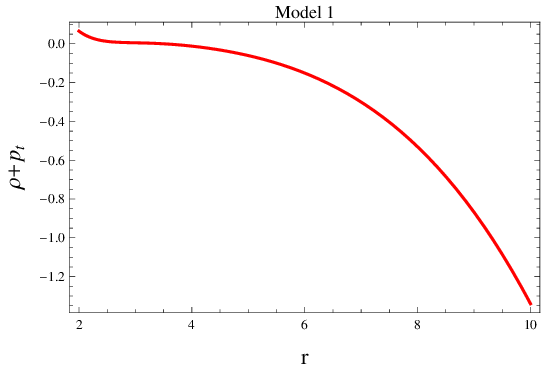,width=.5\linewidth}
\epsfig{file=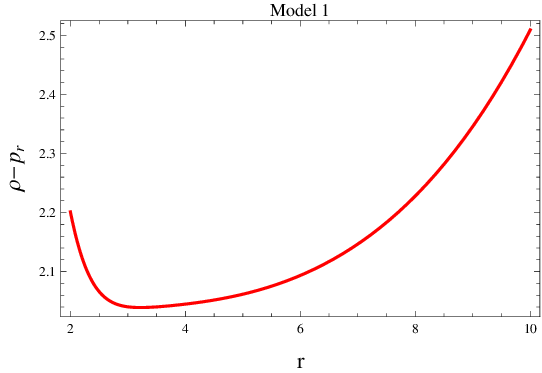,width=.5\linewidth}
\epsfig{file=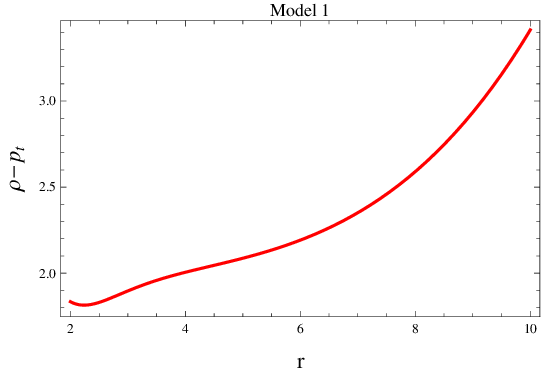,width=.5\linewidth}
\epsfig{file=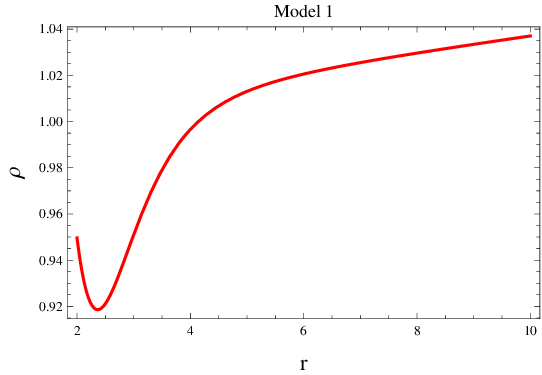,width=.5\linewidth}
\epsfig{file=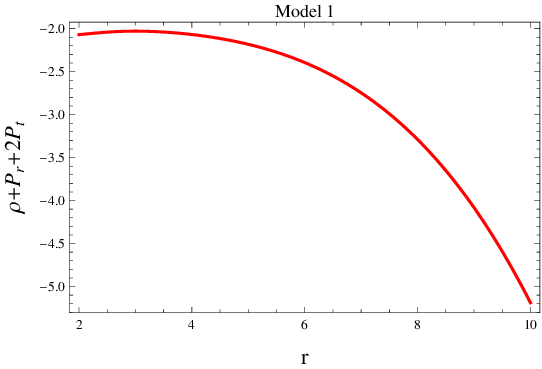,width=.5\linewidth}\caption{Plots of energy
conditions for $\gamma_{1}=-2$, $\xi_{1}=-0.0003$, $m_{1}=-0.0005$.}
\end{figure}

Figure \textbf{6} manifests the behavior of energy conditions for
positive values of $\gamma_{1}$, $\xi_{1}$ and $m_{1}$. The upper
panel violates the null energy condition because $\rho+p_{r}$ and
$\rho+p_{t}$ show negative behavior. However, the positive behavior
of $\rho-p_{r}$ and $\rho-p_{t}$ satisfies the dominant energy
condition. We also note that the $\rho$ is positive but the negative
behavior of $\rho+p_{r}$ and $\rho+p_{t}$ violates the weak energy
condition. The representation of $\rho+p_{r}+2p_{t}$ exhibits
negative trend as shown in the below panel which implies that the
strong energy condition is also violated. These graphs manifest that
fluid parameters violate the energy conditions especially the
violation of the null energy condition for both positive as well as
negative values of $\gamma_{1}$, $\xi_{1}$ and $m_{1}$ provide
viable traversable WH structure in $f(\mathcal{G})$ gravity model.
However, we have also checked that the energy bounds violate for
alternative parametric values.

\subsection{Model 2}

Here, we use another model as \cite{45}
\begin{equation}\nonumber
f(\mathcal{G})=\gamma_{2}\mathcal{G}^{m_{2}}(\xi_{2}\mathcal{G}^{n}+1),
\end{equation}
where $\gamma_{2}$, $\xi_{2}$ and $n$ are arbitrary constant and
$m_{2}>0$. This model is extremely useful for the dealing with
finite time future singularities. The corresponding field equations
are
\begin{eqnarray}\nonumber
\rho&=&\frac{e^{-2b}}{2r^{2}}\bigg[-2e^{b}+2e^{2b}-e^{2b}r^{2}
(\gamma_{2}\mathcal{G}^{m_{2}}(\xi_{2}\mathcal{G}^{n}+1))+e^{2b}r^{2}
\mathcal{G}\big\{\gamma_{2}\xi_{2}(m_{2}+n)
\\\nonumber
&\times&\mathcal{G}^{m_{2}+n-1}+\gamma_{2}m_{2}\mathcal{G}^{m_{2}-1}\big\}
+2b'\bigg\{re^{b}-2(e^{b}-3)\big\{(m_{2}+n)(m_{2}+n-1)
\\\nonumber
&\times&
\gamma_{2}\xi_{2}\mathcal{G}'\mathcal{G}^{m_{2}+n-2}+\gamma_{2}m_{2}
(m_{2}-1)\mathcal{G}'\mathcal{G}^{m_{2}-2}\big\}\bigg\}
-8(1-e^{b})\bigg[(m_{2}+n-1)
\\\nonumber
&\times&
\gamma_{2}\xi_{2}(m_{2}+n)\mathcal{G}''\mathcal{G}^{m_{2}+n-2}
+\gamma_{2}m_{2}(m_{2}-1)\mathcal{G}''\mathcal{G}^{m_{2}-2}
+\gamma_{2}\xi_{2}(m_{2}+n-1)
\\\nonumber
&\times&(m_{2}+n)(m_{2}+n-2)\mathcal{G}'^{2}\mathcal{G}^{m_{2}+n-3}
+\gamma_{2}m_{2}(m_{2}-1)(m_{2}-2)
\mathcal{G}'^{2}\mathcal{G}^{m_{2}-3}\bigg]
\\\nonumber
&-&\frac{2q^{2}r^{2}}{8\pi r^{4}e^{-2b}}\bigg],
\\\nonumber
p_{r}&=&\frac{e^{-2b}}{2r^{2}}\bigg[e^{b}(2+e^{b}(r^{2}(\gamma_{2}
\mathcal{G}^{m_{2}}(\xi_{2}\mathcal{G}^{n}+1))-2))-e^{2b}r^{2}
\mathcal{G}\big\{\gamma_{2}\xi_{2}\mathcal{G}^{m_{2}+n-1}
\\\nonumber
&\times&(m_{2}+n)+\gamma_{2}m_{2}\mathcal{G}^{m_{2}-1}\big\}
+2a'\bigg\{re^{b}-2(e^{b}-3)\big\{(m_{2}+n)(m_{2}+n-1)
\\\nonumber
&\times&\gamma_{2}\xi_{2}\mathcal{G}'\mathcal{G}^{m_{2}+n-2}
+\gamma_{2}m_{2}(m_{2}-1)\mathcal{G}'\mathcal{G}^{m_{2}-2}
\big\}\bigg\}
+\frac{2q^{2}r^{2}}{8\pi r^{4}e^{-2b}}\bigg],
\\\nonumber
p_{t}&=&\frac{e^{-2b}}{4r}\bigg[-2e^{2b}r
\mathcal{G}f_{\mathcal{G}}+a'^{2}(re^{b}+4\mathcal{G}'
f_{\mathcal{G}\mathcal{G}})
+2\bigg\{e^{2b}r(\gamma_{2}\mathcal{G}^{m_{2}}(\xi_{2}\mathcal{G}^{n}+1))
\\\nonumber
&-&b'e^{b}+\big\{e^{b}r+4\big\{\gamma_{2}\xi_{2}(m_{2}+n)(m_{2}+n-1)
\mathcal{G}'\mathcal{G}^{m_{2}+n-2}+\gamma_{2}m_{2}(m_{2}-1)
\\\nonumber
&\times&
\mathcal{G}'\mathcal{G}^{m_{2}-2}\big\}\big\}a''\bigg\}
+a'\bigg[-b'\big\{e^{b}r+12\big\{\gamma_{2}\xi_{2}(m_{2}+n)(m_{2}+n-1)
\mathcal{G}'\mathcal{G}^{m_{2}+n-2}
\\\nonumber
&+&\gamma_{2}m_{2}(m_{2}-1)\mathcal{G}'\mathcal{G}^{m_{2}-2}\big\}\big\}
+2\bigg\{e^{b}+4\big\{\gamma_{2}\xi_{2}(m_{2}+n)(m_{2}+n-1)
\mathcal{G}''\mathcal{G}^{m_{2}+n-2}
\\\nonumber
&+&\gamma_{2}m_{2}(m_{2}-1)\mathcal{G}''\mathcal{G}^{m_{2}-2}
+\gamma_{2}\xi_{2}(m_{2}+n)(m_{2}+n-1)(m_{2}+n-2)\mathcal{G}'^{2}
\mathcal{G}^{m_{2}+n-3}
\\\nonumber
&+&\gamma_{2}m_{2}(m_{2}-1)(m_{2}-2)\mathcal{G}'^{2}
\mathcal{G}^{m_{2}-3}\big\}\bigg\}\bigg]-\frac{2q^{2}r^{2}}{8\pi
r^{4}e^{-2b}}\bigg].
\end{eqnarray}
We consider $m_{2}=0.0005$ for our convenience in the graphical
analysis. We investigate the viable characteristics of WH by
analyzing the above equations. Figure \textbf{7} determines the
behavior of energy bounds for positive values of $\gamma_{2}$,
$\xi_{2}$ and $n$. In the upper panel, the negative behavior of
$\rho+p_{r}$ and $\rho+p_{t}$ show that the null energy condition is
violated. Furthermore, the graphs in the middle part manifest that
the dominant energy condition is satisfied as the behavior of $\rho
- p_r$ and $\rho - p_t$ is positive. The components $\rho$ and $\rho
+ p_r + 2p_t$ also exhibit negative trends, indicating the violation
of strong and weak energy conditions, respectively. Figure
\textbf{8} represents the energy conditions for negative values of
the model parameters. These graphs show that the fluid parameters
satisfy the energy conditions, i.e., the behavior of matter
components $(\rho, \rho\pm p_{r}, \rho\pm p_{t})$ is positive for
negative values of $\gamma_{2}$, $\xi_{2}$ and $n$ which yields
non-traversable WH structure. Hence, in this gravity model, a viable
traversable WH structure can be obtained for positive values of the
model parameter.
\begin{figure}
\epsfig{file=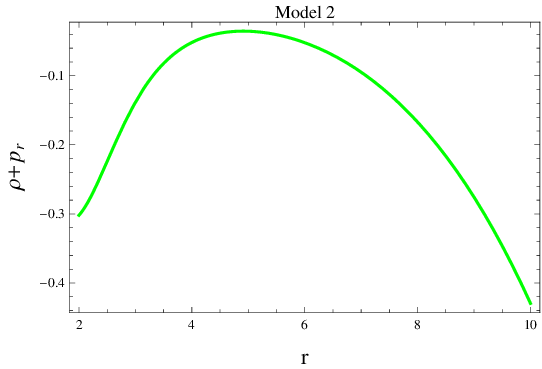,width=.5\linewidth}
\epsfig{file=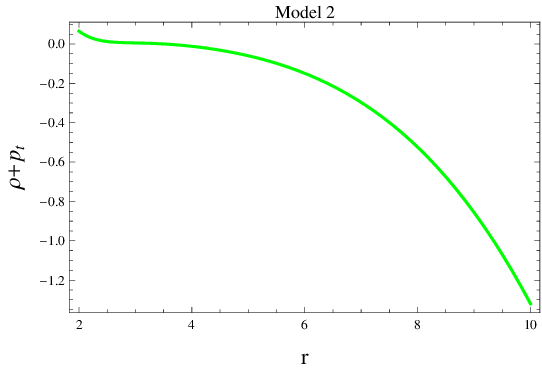,width=.5\linewidth}
\epsfig{file=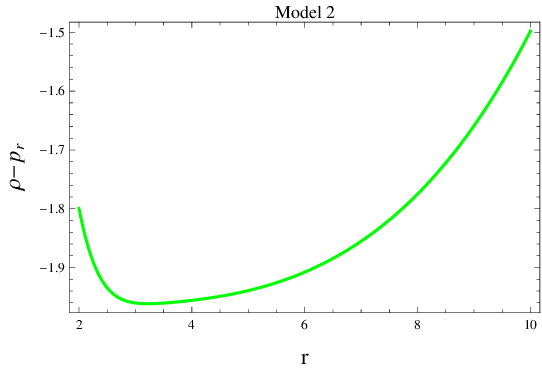,width=.5\linewidth}
\epsfig{file=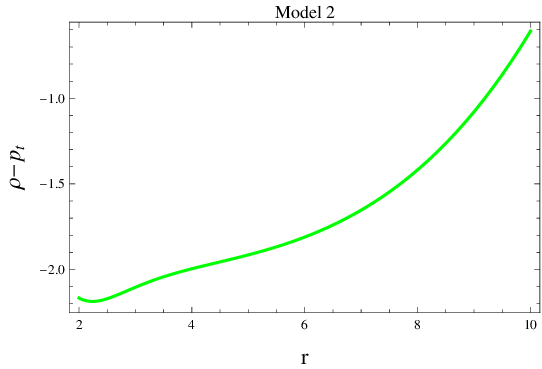,width=.5\linewidth}
\epsfig{file=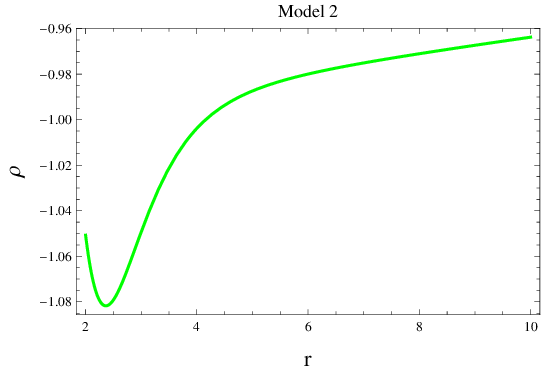,width=.5\linewidth}
\epsfig{file=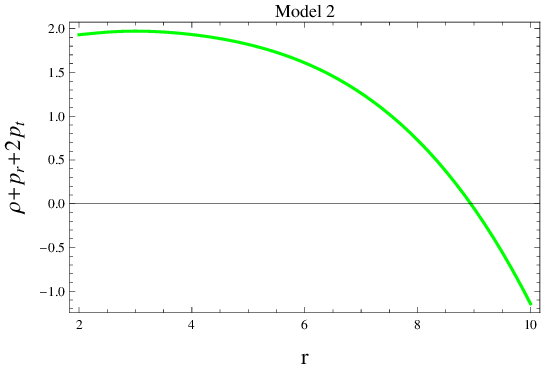,width=.5\linewidth}\caption{Plots of energy
conditions for $\gamma_{2}=2$, $\xi_{2}=0.0003$ and $n=0.001$.}
\end{figure}
\begin{figure}
\epsfig{file=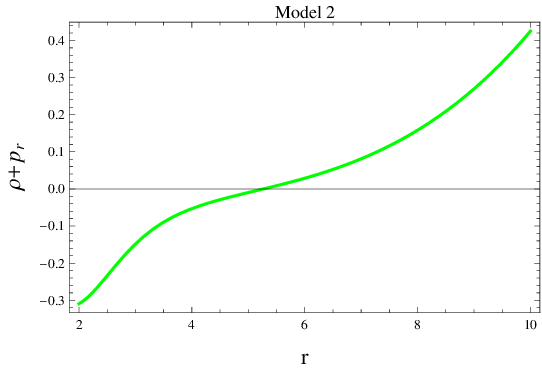,width=.5\linewidth}
\epsfig{file=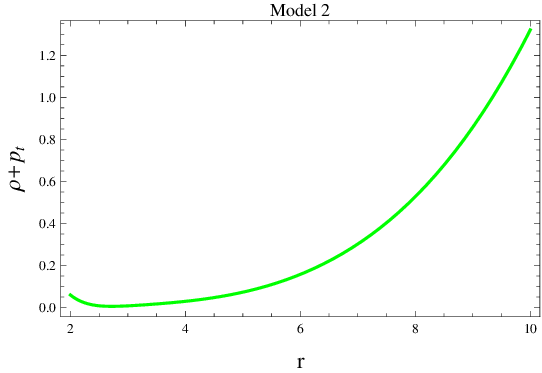,width=.5\linewidth}
\epsfig{file=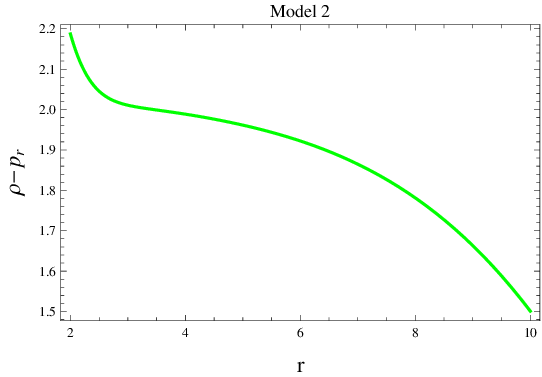,width=.5\linewidth}
\epsfig{file=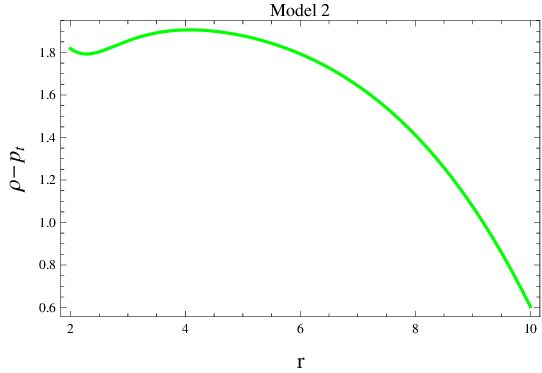,width=.5\linewidth}
\epsfig{file=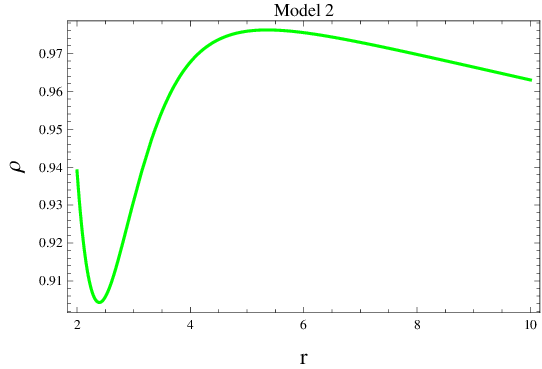,width=.5\linewidth}
\epsfig{file=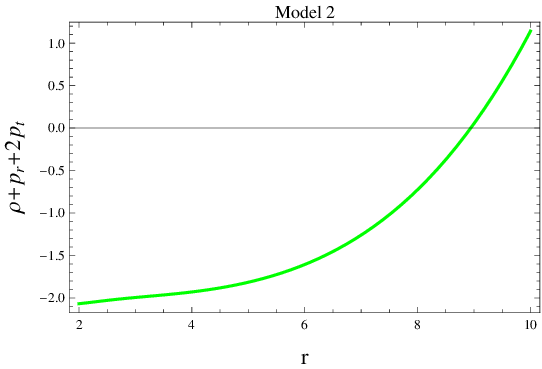,width=.5\linewidth}\caption{Plots of energy
conditions for $\gamma_{2}=-2$, $\xi_{2}=-0.0003$ and $n=-0.001$.}
\end{figure}

\subsection{Model 3}

Finally, we consider the following viable model as
\begin{equation}\nonumber
f(\mathcal{G})=\frac{\gamma_{3}\mathcal{G}^{m_{3}}+\xi_{3}}
{\gamma_{4}\mathcal{G}^{m_{3}}+\xi_{4}}.
\end{equation}
Here $\gamma_{3}$, $\gamma_{4}$, $\xi_{3}$ and $\xi_{4}$ are
arbitrary constants with $m_{3}>0$. The corresponding field
equations are
\begin{eqnarray}\nonumber
\rho&=&\frac{e^{-2b}}{2r^{2}}\bigg[-2e^{b}+2e^{2b}-e^{2b}
r^{2}(\gamma_{3}\mathcal{G}^{m_{3}}
+\xi_{3})(\gamma_{4}\mathcal{G}^{m_{3}}+\xi_{4})^{-1}+
\frac{e^{2b}r^{2}}{(\gamma_{4}\mathcal{G}^{m_{3}}+\xi_{4})^{2}}
\\\nonumber
&\times&\big\{(\gamma_{4}\mathcal{G}^{m_{3}}+\xi_{4})
(\gamma_{3}m_{3}\mathcal{G}^{m_{3}-1})
+(\gamma_{3}\mathcal{G}^{m_{3}}+\xi_{3})(\gamma_{4}
m_{3}\mathcal{G}^{m_{3}-1})\big\}
+2b'\bigg[re^{b}
\\\nonumber
&-&\frac{2(e^{b}-3)\mathcal{G}'}{(
\gamma_{4}\mathcal{G}^{m_{3}}+\xi_{4})^{3}}\bigg[
m_{3}\bigg\{(\mathcal{G}^{2m_{3}-2}\xi_{3}
\gamma_{4}^{2}m_{3}-\mathcal{G}^{2m_{3}-2}\xi_{4}\gamma_{3}
\gamma_{4}m_{3}+\mathcal{G}^{2m_{3}-2}\xi_{3}
\gamma_{4}^{2}-\xi_{4}\gamma_{3}
\\\nonumber
&\times&\gamma_{4}\mathcal{G}
^{2m_{3}-2}-\mathcal{G}^{m_{3}-2}\xi_{3}\xi_{4}
\gamma_{4}m_{3}+\mathcal{G}^{m_{3}-2}\xi_{4}^{2}
\gamma_{3}m_{3}+\mathcal{G}^{m_{3}-2}
\xi_{3}\xi_{4}\gamma_{4}-\mathcal{G}^{m_{3}-2}
\xi_{4}^{2}\gamma_{3}\bigg\}\bigg]\bigg]
\\\nonumber
&-&8\mathcal{A}(1-e^{b})-\frac{2q^{2}r^{2}}{8\pi
r^{4}e^{-2b}}\bigg],
\\\nonumber
p_{r}&=&\frac{e^{-2b}}{2r^{2}}\bigg[e^{b}(2+e^{b}
\big\{(r^{2}(\gamma_{3}\mathcal{G}^{m_{3}}
+\xi_{3})(\gamma_{4}\mathcal{G}^{m_{3}}
+\xi_{4})^{-1}-2)\big\}-
\frac{e^{2b}r^{2}\mathcal{G}}{(\gamma_{4}\mathcal{G}^{m_{3}}
+\xi_{4})^{2}}
\\\nonumber
&\times&
\big\{(\gamma_{4}\mathcal{G}^{m_{3}}+\xi_{4})(\gamma_{3}m_{3}
\mathcal{G}^{m_{3}-1})
+(\gamma_{3}\mathcal{G}^{m_{3}}+\xi_{3})(\gamma_{4}m_{3}
\mathcal{G}^{m_{3}-1})\big\}
+2a'\bigg\{re^{b}
\\\nonumber
&-&\frac{2(e^{b}-3)\mathcal{G}'}{(\gamma_{4}\mathcal{G}^{m_{3}}
+\xi_{4})^{2}}
\big\{(\gamma_{4}\mathcal{G}^{m_{3}}+\xi_{4})(\gamma_{3}m_{3}
\mathcal{G}^{m_{3}-1})
+(\gamma_{3}\mathcal{G}^{m_{3}}+\xi_{3})(\gamma_{4}m_{3}
\mathcal{G}^{m_{3}-1})\big\}\bigg\}
\\\nonumber
&+&\frac{2q^{2}r^{2}}{8\pi r^{4}e^{-2b}}\bigg],
\\\nonumber
p_{t}&=&\frac{e^{-2b}}{4r}\bigg[-
\frac{2e^{2b}r\mathcal{G}}{(\gamma_{4}\mathcal{G}^{m_{3}}
+\xi_{4})^{2}}
\big\{(\gamma_{4}\mathcal{G}^{m_{3}}+\xi_{4})(\gamma_{3}
m_{3}\mathcal{G}^{m_{3}-1})
+(\gamma_{4}m_{3}\mathcal{G}^{m_{3}-1})
\\\nonumber
&\times&(\gamma_{3}\mathcal{G}^{m_{3}}+\xi_{3})\big\}
+a'^{2}\bigg\{re^{b}+\frac{4\mathcal{G}'}{(
\gamma_{4}\mathcal{G}^{m_{3}}+\xi_{4})^{3}}\big\{
m_{3}\big\{\mathcal{G}^{2m_{3}-2}\xi_{3}
\gamma_{4}^{2}m_{3}-\xi_{4}\gamma_{3} \gamma_{4}
\\\nonumber
&\times&m_{3}\mathcal{G}^{2m_{3}-2}+\mathcal{G}^{2m_{3}-2}
\xi_{3}\gamma_{4}^{2}-\mathcal{G}
^{2m_{3}-2}b_{4}\gamma{3}\gamma_{4}-\mathcal{G}^{m_{3}-2}
\xi_{3}\xi_{4}
\gamma_{4}m_{3}+\mathcal{G}^{m_{3}-2}\xi_{4}^{2}\gamma_{3}m_{3}
\\\nonumber
&+&\mathcal{G}^{m_{3}-2}
\xi_{3}\xi_{4}\gamma_{4}-\mathcal{G}^{m_{3}-2}\xi_{4}^{2}
\gamma_{3}\big\}\big\}\bigg\}
+2\bigg[e^{2b}r(\gamma_{3}\mathcal{G}^{m_{3}}
+\xi_{3})(\gamma_{4}\mathcal{G}^{m_{3}}+\xi_{4})^{-1} -b'e^{b}
\\\nonumber
&+&\bigg\{e^{b}r+\frac{4\mathcal{G}'}{(
\gamma_{4}\mathcal{G}^{m_{3}}+\xi_{4})^{3}}\big\{
m_{3}\big\{\mathcal{G}^{2m_{3}-2}\xi_{3}\gamma_{4}^{2}m_{3}
-\mathcal{G}^{2m_{3}-2}\xi_{4}\gamma_{3}
\gamma_{4}m_{3} +\mathcal{G}^{2m_{3}-2}\xi_{3}\gamma_{4}^{2}
\\\nonumber
&-&\mathcal{G}
^{2m_{3}-2}\xi_{4}\gamma_{3}\gamma_{4}-\mathcal{G}^{m_{3}-2}
\xi_{3}\xi_{4}
\gamma_{4}m_{3}+\mathcal{G}^{m_{3}-2}\xi_{4}^{2}\gamma_{3}
m_{3}+\mathcal{G}^{m_{3}-2}
\xi_{3}\xi_{4}\gamma_{4}-\mathcal{G}^{m_{3}-2}\xi_{4}^{2}
\gamma_{3}
\\\nonumber
&\times&\big\}\big\}\bigg\}a''\bigg]
+a'\bigg[-b'\bigg\{e^{b}r+\frac{12\mathcal{G}'}{(
\gamma_{4}\mathcal{G}^{m_{3}}+\xi_{4})^{3}} \big\{
m_{3}\big\{\mathcal{G}^{2m_{3}-2}\xi_{3}
\gamma_{4}^{2}m_{3}-\mathcal{G}^{2m_{3}-2}\xi_{4}\gamma_{3}
\gamma_{4}m_{3}
\\\nonumber
&+&\mathcal{G}^{2m_{3}-2}\xi_{3}\gamma_{4}^{2}-\mathcal{G}
^{2m_{3}-2}\xi_{4}\gamma_{3}\gamma_{4}-\mathcal{G}^{m_{3}-2}
\xi_{3}\xi_{4}
\gamma_{4}m_{3}+\mathcal{G}^{m_{3}-2}\xi_{4}^{2}\gamma_{3}
m_{3}+\mathcal{G}^{m_{3}-2}
\xi_{3}\xi_{4}\gamma_{4}
\\\nonumber
&-&\mathcal{G}^{m_{3}-2}\xi_{4}^{2}\gamma_{3}\big\}\big\}
\bigg\}
+2(e^{b}+4\mathcal{A})\bigg]-\frac{2q^{2}r^{2}}{8\pi
r^{4}e^{-2b}}\bigg].
\end{eqnarray}
where
\begin{eqnarray}\nonumber
\mathcal{A}&=&\frac{\mathcal{G}''}{(
\gamma_{4}\mathcal{G}^{m_{3}}+\xi_{4})^{3}}\bigg[
m_{3}\bigg\{(\mathcal{G}^{2m_{3}-2}\xi_{3}
\gamma_{4}^{2}m_{3}-\mathcal{G}^{2m_{3}-2}\xi_{4}
\gamma_{3}
\gamma_{4}m_{3}
+\mathcal{G}^{2m_{3}-2}\xi_{3}\gamma_{4}^{2}-\xi_{4}
\gamma_{3}\gamma_{4}
\\\nonumber
&\times&\mathcal{G} ^{2m_{3}-2}-\mathcal{G}^{m_{3}-2}
\xi_{3}\xi_{4}
\gamma_{4}m_{3}+\mathcal{G}^{m_{3}-2}\xi_{4}^{2}
\gamma_{3}m_{3}+\mathcal{G}^{m_{3}-2}
\xi_{3}\xi_{4}\gamma_{4}-\mathcal{G}^{m_{3}-2}
\xi_{4}^{2}\gamma_{3}\bigg\}\bigg]
\\\nonumber
&+&\frac{\mathcal{G}'^{2}}{(\gamma_{4}
\mathcal{G}^m_{3}+\xi_{4}
)^{4}}\bigg[-m_{3}\bigg\{
\mathcal{G}^{m_{3}-3}\xi_{3}\xi_{4}^{2}\gamma_{4}m_{3}^{2}-
\mathcal{G}^{m_{3}-3}\xi_{4}^{3}\gamma_{3}m_{3}^{2}
+\mathcal{G}^{3m_{3}-3}b_{3}\gamma_{4}^{3}
m_{3}^{2}
\\\nonumber
&-&\mathcal{G}^{3m_{3}-3}\xi_{4}\gamma_{3}
\gamma_{4}^{2}m_{3}^{2}-4
\mathcal{G}^{2m_{3}-3}\xi_{3}\xi_{4}\gamma_{4}^{2}m_{3}^{2}
+4\mathcal{G}^{2m_{3}-3
}\xi_{4}^{2}\gamma_{3}\gamma_{4}m_{3}^{2}-3\mathcal{G}^{m_{3}
-3}\xi_{3}\xi_{4}^{2}
\gamma_{4}m_{3}
\\\nonumber
&+&3\mathcal{G}^{m_{3}-3}\xi_{4}^{3}\gamma_{3}m+3
\gamma_{4}^{3}m_{3}\mathcal{G}^{3m_{3}-3}\xi_{3}-3
\mathcal{G}^{3m_{3}-3}\xi_{4}\gamma_{3}
\gamma_{4}^{2}m_{3}+2\mathcal{G}^{m_{3}-3}\xi_{3}
\xi_{4}^{2}\gamma_{4}
\\\nonumber
&-&2\mathcal{G}^{m_{3}-
3}\xi_{4}^{3}\gamma_{3}+2\mathcal{G}^{3m_{3}-3}\xi_{3}
\gamma_{4}^{3}-2\mathcal{G}
^{3m_{3}-3}\xi_{4}\gamma_{3}\gamma_{4}^{2}
+4\mathcal{G}^{2m_{3}-3}b_{3}b_{4}
\gamma_{4}^{2}
\\\nonumber
&-&4\mathcal{G}^{2m_{3}-3}\xi_{4}^{2}
\gamma_{3}\gamma_{4}\bigg\}\bigg].
\end{eqnarray}
Figures \textbf{9} and \textbf{10} show the behavior of energy
conditions for positive and negative values of $\gamma_{3}$,
$\gamma_{4}$, $\delta_3$ and $\delta_4$. The graphs reveal that the
matter components $(\rho, \rho\pm p_{r}, \rho\pm p_{t})$ show
negative behavior for all parametric values. This violation of
energy conditions indicate the presence of exotic matter, which
justifies the existence of a viable traversable WH geometry in this
gravity model.
\begin{figure}
\epsfig{file=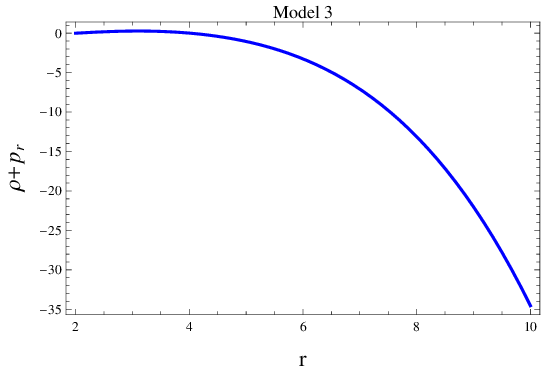,width=.5\linewidth}
\epsfig{file=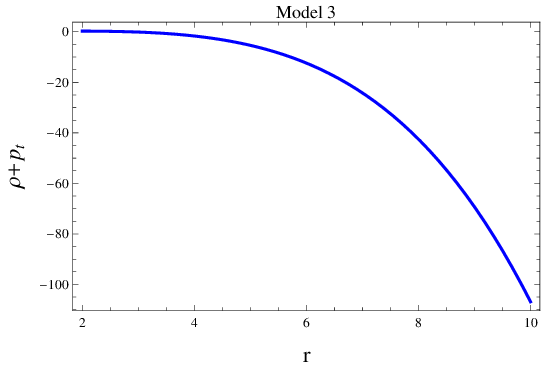,width=.5\linewidth}
\epsfig{file=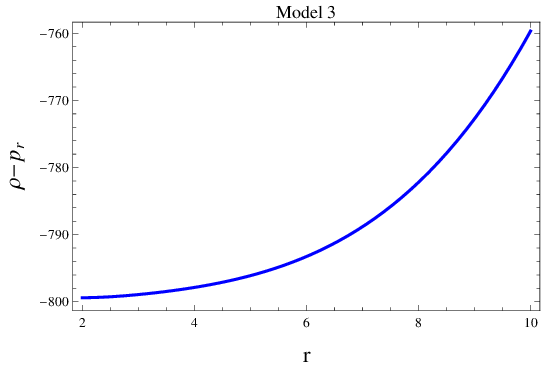,width=.5\linewidth}
\epsfig{file=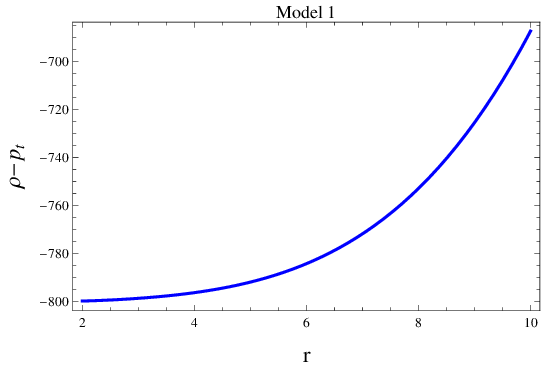,width=.5\linewidth}
\epsfig{file=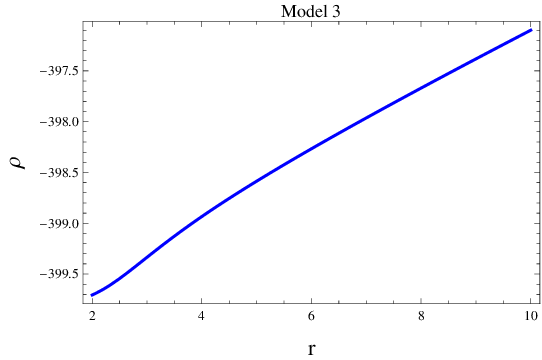,width=.5\linewidth}
\epsfig{file=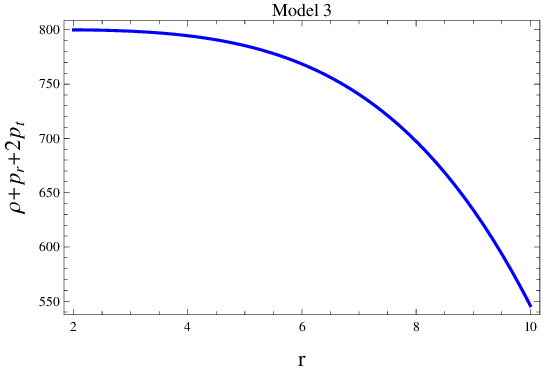,width=.5\linewidth}\caption{Plots of energy
conditions for $\gamma_{3}=2$, $\gamma_{4}=0.002$, $\xi_{3}=0.0003$
$\xi_{4}=0.0005$ and $n=0.001$.}
\end{figure}
\begin{figure}
\epsfig{file=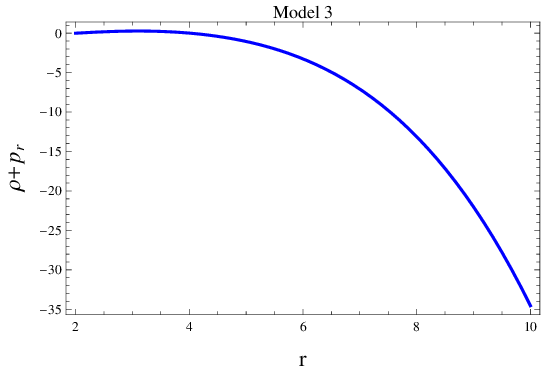,width=.5\linewidth}
\epsfig{file=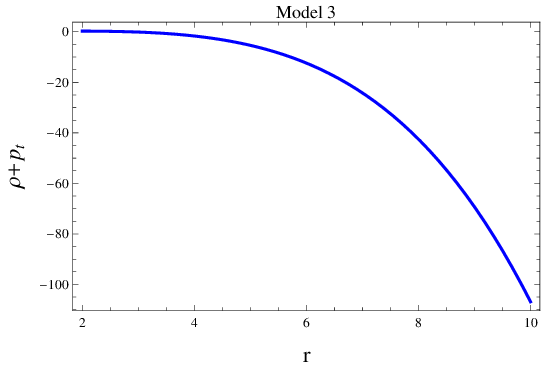,width=.5\linewidth}
\epsfig{file=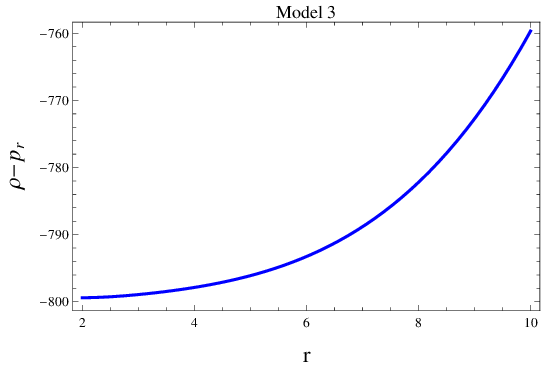,width=.5\linewidth}
\epsfig{file=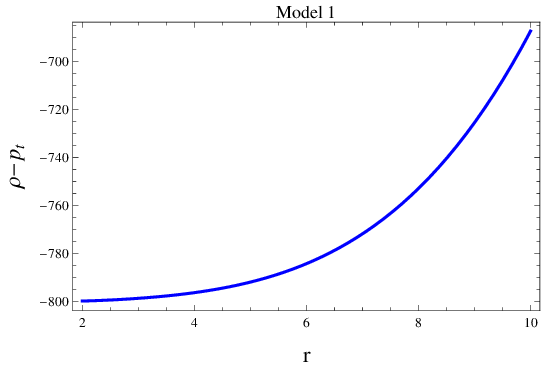,width=.5\linewidth}
\epsfig{file=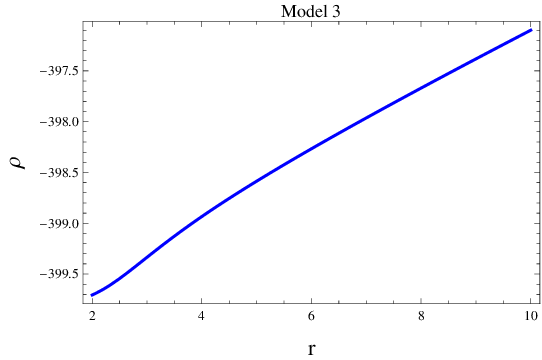,width=.5\linewidth}
\epsfig{file=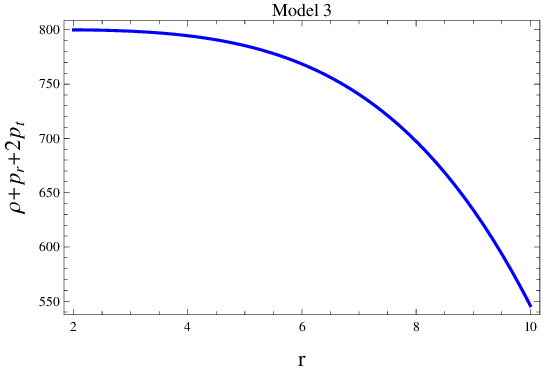,width=.5\linewidth}\caption{Plots of energy
conditions for $\gamma_{3}=-2$, $\gamma_{4}=-0.002$,
$\xi_{3}=-0.0003$ $\xi_{4}=-0.0005$ and $n=0.001$}
\end{figure}

\section{Concluding Remarks}

Various methods have been used in literature to obtain viable WH
structures. One of them is to formulate shape function through
different methods and other is to examine the behavior of energy
constraints by considering different WSFs. The energy conditions can
be used to derive important physical properties of the object. In
the present article, we have studied the viable WH geometry through
embedding class-I in $f(\mathcal{G})$ gravity. In this perspective,
we have built a shape function by employing the Karmarkar condition
to check whether WH solutions exist or not in this theory. We have
considered three different models of this modified theory to find
the exact solutions of static spherical spacetime. We have examined
the viability of traversable WH geometry through the energy
conditions. The obtained results are summarized as follows
\begin{itemize}
\item
The newly developed shape function through the Karmarkar condition
satisfies all the necessary conditions which ensure the presence of
physically viable WH geometry (Figure \textbf{1}).
\item
we have discussed embedded diagram to represent the WH structure. We
have considered equatorial slice $\theta=\frac{\pi}{2}$ and a fixed
moment of time i.e., $t$=constant for spherical symmetry and for the
visualization, we embed it into three dimensional Euclidean space.
Moreover, one can visualize the upper universe for $h>0$ and the
lower universe $h<0$ (Figures \textbf{2} and \textbf{3}).
\item
Figure \textbf{4} shows the graphical behavior of energy conditions
for $f(\mathcal{G})=0$. The matter components $(\rho, \rho+p_{r},
\rho-p_t, \rho+p_r+2p_t)$ show negative behavior, indicating the
violation of weak, null, dominant and strong energy conditions,
respectively. Thus, a viable traversable WH structure can be
obtained due to the presence of exotic matter in this gravity model.
\item
For the first model, we have shown that the fluid parameters violate
the energy conditions especially the violation of the null energy
condition for both positive/negative values of $\gamma_{1}$,
$\xi_{1}$ and $m_{1}$ which gives the existence of exotic matter at
the WH throat (Figures \textbf{5} and \textbf{6}). Thus, we have
obtained the viable traversable WH geometry for all parametric
values.
\item
For the second model, we have obtained viable traversable WH
structure for positive values of model parameters, because the
behavior of matter components $(\rho, \rho\pm p_{r}, \rho\pm p_{t})$
is negative which ensures the presence of exotic matter at the WH
throat (Figure \textbf{7}). But, for negative values of the model
parameters, we have obtained non-traversable WH geometry as the
energy conditions are satisfied which gives the existence of normal
matter at the WH throat (Figure \textbf{6}).
\item
The energy conditions are satisfied for both positive as well as
negative values of model parameters, which show that the viable
traversable WH geometry exists for the third $f(\mathcal{G})$ model
(Figures \textbf{9} and \textbf{10}).
\end{itemize}
Shamir and Fayyaz \cite{46} examined physically viable WH structure
via Karmarkar condition in $f(\mathfrak{R})$ theory and obtained
viable WH solutions in the presence of minimum amount of exotic
matter. Recently, Sharif and Fatima \cite{41} generalized this work
for $f(\mathfrak{R},\mathcal{T})$ theory and obtained viable WH
solutions for minimum values of radius. It is noteworthy to mention
here that we have found viable WH solutions in $f(\mathcal{G})$
gravity as energy conditions are violated which gives the existence
of exotic matter at WH throat. Our investigation has explored
traversable WH solutions by incorporating energy conditions, which
are composed of energy density and pressure components, encompassing
the Gauss-Bonnet terms. Consequently, the WH constructions presented
in this manuscript are well established. We conclude that physically
viable traversable WH solutions through
Karmarkar condition exist in $f(\mathcal{G})$ theory.\\\\
\textbf{Data Availability Statement:} No new data were created or
analyzed in this study.

\end{document}